\documentclass[aps,prc,preprint,showpacs,preprintnumbers,nofootinbib,float,superscriptaddress,longbibliography]{revtex4-1}
\usepackage{graphicx, fancybox}
\usepackage{amsmath,amssymb}
\usepackage[colorlinks=true, pdfstartview=FitV, linkcolor=red,
		     citecolor=blue, urlcolor=blue]{hyperref}
\usepackage{color}
\usepackage{soul}
\usepackage{url}
\usepackage[normalem]{ulem}
\usepackage{nccmath}
\usepackage{amsthm}
\usepackage{mathrsfs}
\usepackage{graphicx}
\usepackage{caption}
\usepackage{subcaption}
\usepackage{float}

\begin{document}
   
\title{\bf Holographic spin liquids and Lovelock Chern-Simons gravity}
\author{A.D. Gallegos$^1$ and U. G\"ursoy}
\affiliation{Institute for Theoretical Physics, Utrecht University, Leuvenlaan 4, 3584 CE Utrecht, The Netherlands}
\date{\today}


\begin{abstract}
We explore the role of torsion as source of spin current in strongly interacting conformal fluids using holography.  We establish the constitutive relations of the basic hydrodynamic variables, the energy-momentum tensor and the spin current based on the classification of the spin sources in irreducible Lorentz representations. The fluids we consider are assumed to be described by the five dimensional Lovelock-Chern-Simons gravity with independent vielbein and spin connection. We construct a hydrodynamic expansion that involves the stress tensor and the spin current and compute the corresponding one-point functions holographically. As a byproduct we find a class of interesting analytic solutions to the Lovelock-Chern-Simons gravity, including blackholes, by mapping the equations of motion into non-linear algebraic constraints for the sources. We also derive a Lee-Wald entropy formula for these blackholes in Chern-Simons theories with torsion. The blackhole solutions determine the thermodynamic potentials and the hydrodynamic constitutive relations in the corresponding fluid on the boundary. We observe novel spin induced transport in these holographic models: a dynamical version of the Barnett effect where vorticity generates a spin current and anomalous vortical transport transverse to a vector-like spin source.

\end{abstract}
\maketitle

\newpage

\section{Introduction}\label{sec::intro}

Hydrodynamic description of strongly correlated systems with spin degrees of freedom is an important open problem with various applications ranging from condensed matter to astrophysics and high energy physics \cite{Baier:2006um,Huovinen:2006jp,Gale:2013da,Heinz:2013th,deSouza:2015ena}. In the relativistic limit, it is particularly relevant for the quark-gluon plasma which is produced along with strong magnetic fields \cite{Kharzeev:2007jp,Skokov:2009qp,Tuchin:2010vs,Voronyuk:2011jd,Deng:2012pc,Tuchin:2013ie,McLerran:2013hla,Gursoy:2014aka} which presumably magnetizes the plasma and produces macroscopic flow of spin degrees of freedom \cite{Kharzeev:2007jp,Fukushima:2008xe,Kharzeev:2010gd,Burnier:2011bf}. Fluid-like description of spin dynamics in non-relativistic systems are also crucial in spintronics \cite{Bandurin1055,Crossno1058} and quantum spin liquids \cite{Savary:2016ksw}. We are interested in relativistic systems in this paper which include not only the quark-gluon plasma but also the fluid-like phases with relativistic dispersion which arise in condensed matter. 

Relativistic hydrodynamics with spin degrees of freedom is ambiguous. The spin current inherits the ambiguity in the definition of the energy-momentum tensor. In particular, the total angular momentum can be written as sum of orbital and spin components as 
\begin{align}
\begin{split}\label{decompositionJ}
J^{\lambda \mu \nu} &= x^\mu T^{\lambda \nu} - x^{\nu} T^{\lambda \mu} +S^{\lambda \mu \nu}\, ,
\end{split}
\end{align}
where $T^{\mu \nu}$ is the canonical energy momentum tensor obtained from Noether's theorem as the charge under space-time translations and $S^{\lambda \mu \nu}$ is the relativistic generalization of $S^{ijk}$ i.e. the current in the ith direction of spin orthogonal to the $j,k$ plane. Even though the latter is supposedly present whenever the quantum fields transform non-trivially under rotations, its value changes under the following pseudo-gauge transformation \cite{Hehl}
\begin{align}
\label{psg}
\begin{split}
T^{'\mu \nu} &= T^{\mu \nu} + \frac{1}{2} \nabla_\lambda \left(\Phi^{\lambda \mu \nu} - \Phi^{\mu \lambda \nu}- \Phi^{\nu \lambda \mu} \right) \, , \\
S^{' \lambda \mu \nu} &= S^{\lambda \mu \nu} - \Phi^{\lambda \mu \nu} \, .
\end{split}
\end{align}
where $\Phi^{\lambda \mu \nu}$ is a tensor antisymmetric in the last two indices. This transformation preserves  local conservation of both currents 
\begin{align}
	\partial_\mu T^{\mu \nu} &= 0 \, ,\\ 
	\partial_\lambda J^{\lambda \mu \nu}&= 0\, ,
\end{align}
and, as well understood, is important to render the canonical energy-momentum tensor symmetric by removing its antisymmetric part. In fact, the Belinfante tensor \cite{Belinfante,Rosenfeld} --- which corresponds to the choice $\Phi^{\lambda \mu \nu} = S^{\lambda \mu \nu}$ hence removes the spin current completely --- yields the energy-momentum tensor  that coincides with its definition in gravity\footnote{This is easily derived by coupling field theory to (a priori) independent spin-connection and vielbein, and then requiring metric compatibility and vanishing torsion.}. 
 
These conservation equations together yield 
\begin{align}
\begin{split}
\partial_\lambda S^{\lambda \mu \nu}= -2 T^{[\mu \nu]} \, ,
\end{split}
\end{align}
%
which shows that the spin current would be sourced by the antisymmetric part of the energy-momentum tensor. If this part is removed then one expects a vanishing spin current as in the Belinfante gauge. This conclusion is challenged recently \cite{Becattini1} (see also \cite{Becattini2,Becattini3}) where it was argued that, even though the definition of the spin current operator is subject to the pseudo-gauge transformation (\ref{psg}) this symmetry might be broken by the thermodynamic  state of the theory, hence the different pseudo-gauges would lead to different observables. Becattini et al. argue that this observation may shed light on the recent observation of macroscopic polarization in the quark-gluon plasma \cite{STAR:2017ckg}. 

In this paper, we take a different viewpoint and consider a field theory with non-trivial torsion. Presence of torsion guarantees a non-trivial spin current as it produces a spin connection which in turn sources the spin current \cite{Kibble,Sciama,Obukhov}. This observation has also been made in the condensed-matter literature where geometric torsion is related to lattice defects \cite{TorsionC2,TorsionC1,CT3}. Following this observation relativistic and non-relativistic transport in geometries with torsion have been studied via various approaches \cite{ChiralTorsional,CT2,CT4}. 

We consider a 4D relativistic, strongly interacting field theory with canonical fields that transform non-trivially under rotations and couple this theory to a vierbein $e$ and a spin connection $\omega$ which we take as independent sources. We take the background space-time to have flat metric (flat vierbein) but non-trivial spin-connection which, in Riemann-Cartan geometry, arises from a non-trivial (con)torsion. The hydrodynamics theory in the presence of these sources consists of the two dynamical equations, which generalize the equations for the energy-momentum and spin current above: 
\begin{align}\label{hyd1}
\begin{split}
\nabla_\mu(|e| T^{\mu}_{\nu}) = \frac{|e|}{2} S^{\lambda}_{\hphantom{\lambda} \rho \sigma} R^{\rho \sigma}_{\hphantom{\mu \nu} \lambda \nu } \, , \quad \quad \quad
\nabla_\lambda \left(|e| S^\lambda_{\hphantom{\lambda} \rho \sigma}  \right) = \frac{|e|}{2} T_{[\rho \sigma]} \, ,
\end{split}
\end{align}
where $|e|$ is the determinant of the metric which we set to 1 in flat background, $\nabla$ is the covariant derivative whose connection includes both the Levi-Civita connection and the torsion, and $R$ is the Riemann curvature.  We then construct the hydrodynamic constitutive relations both for $T_{\mu\nu}$ and $S^\lambda_{\hphantom{\lambda} \rho \sigma}$ using space-time symmetries. We use holographic methods \cite{AdS1,AdS2,AdS3}  to solve the hydrodynamic equations and to determine the energy-momentum tensor and the spin current. In particular we use the fluid-gravity correspondence \cite{fluidGravity1,fluidGravity2}, which, in our context, amounts to finding gravitational solutions with torsion order by order in gradient expansion, and using the holographic prescription to compute the currents $\{T^{\mu \nu}, S^{\lambda \mu \nu} \}$ in the dual quantum field theory. 

For this purpose we focus on the 5 dimensional Lovelock-Chern-Simons (LCS) gravity without matter \cite{Chamseddine:1989nu,Zanelli:2005sa,Banados}. We pick this theory two reasons: unlike Einstein gravity (1) the spin connection and vielbein are independent, and  (2) the equations of motion can be reduced to algebraic equations. The first point is crucial for obtaining a non-trivial spin current as discussed above. The second is practical. 

Use of holographic techniques in the context of spin dynamics has a short history. Generalization of standard holography from Einstein gravity to Einstein-Cartan (EC) gravity \cite{Cartan1,Cartan2} where the vielbein and the spin connection become independent started with \cite{Banados} and continued with \cite{Klemm:2007yu,Leigh:2008tt,Petkou:2010ve,Miskovic,Camanho:2013pda,Cvetkovic:2017fxa}. Relevance of spin connection to condensed matter in the holography literature was  considered in \cite{Leigh:2008tt} and \cite{Hashimoto}. 

The paper is organized as follows. In section \ref{sec::QFT} we review how the energy momentum tensor and the spin current in a generic QFT are sourced by the metric and the spin connection, and derive the corresponding conservation equations. In section \ref{sec::hydro} we set up the hydrodynamic limit of the QFT coupled to spin sources, namely we provide the constitutive relations for the energy momentum tensor and the spin current. We classify the spin sources and the one-point functions in terms of irreducible representations of the Lorentz group in 3+1 dimensions. In particular we show that the spin current is decomposed into an axial charge density, a vector charge density, two axial currents, two vector currents, one tensor current, and one pseudo-tensor current, see the table \ref{table}. In the next section, \ref{sec::Holography}, the Chern-Simons model of \cite{Banados} is reviewed and the holographic prescription for obtaining the currents is derived. In section \ref{sec::Hydro} we present an ansatz for the holographic model that is suitable for the fluid-gravity correspondence. We then solve the dynamical equations of motion to all orders in the hydrodynamic derivative expansion by reducing them  to a set of constraint equations. In section \ref{sec::ZerothSolution} we consider zeroth order hydrodynamics, namely constant sources, and present two zeroth order gravitational blackhole solutions. These are  the only solutions with independent and unconstrained spin sources. The energy momentum tensor, spin current, and thermodynamic potentials for the solutions are derived. In section \ref{sec::FirstSolution} we promote all the sources (temperature, four velocity and spin sources) to slowly varying functions of the boundary coordinates and obtain the corresponding first order solutions. We then identify several spin induced transport phenomena by examining these solutions. We summarize our results and provide possible extensions of our work  in section \ref{sec::Conclusions}. 

The three appendices contain crucial details. In appendix A we review the boundary Noether symmetries, the equations of motion for the energy-momentum tensor and the spin current by demanding invariance under these symmetries, and possible anomalies. Appendix B presents a type of irregular blackhole solution which we ignored in our main presentation. Finally, appendix C contains, as far as we know, a novel derivation of the Wald entropy formula for Chern-Simons theories with nontrivial torsion. 
\section{Quantum field theory coupled to first order backgrounds}\label{sec::QFT}

The coupling between spin current and spin connection is best seen in the example of Dirac fermion in four dimensions \cite{KibbleSpin}. To place fermions in curved spacetime it is necessary to change from a formulation in terms of the background metric $\gamma_{\mu \nu}$ to a first order formulation  in terms of the vielbein fields  $e^a=e^a_\mu dx^\mu$ related to the metric by\footnote{We will use the indices $\{A,B,...\}$ for 5D tangent space indices, the indices $\{M,N,...\}$ for 5D spacetime indices, the indices $\{a,b,...\}$ for 4D tangent space indices, the indices $\{ \mu, \nu, ...\}$ for 4D spacetime indices and indices $\{\bar A, \bar B,...\}$ for SO(4,2) indices.} 
\begin{align}
\begin{split}
e^a_\mu e^b_\nu \eta_{ab} &= \gamma_{\mu \nu}.
\end{split}
\end{align}
The vielbein introduce a flat tangent space at every point in spacetime on which the Clifford algebra can be locally defined. This results in the Lagrangian 
\vspace{1mm}
\begin{align}
\begin{split}\label{diracLagrangian}
S[\psi] = \int d^4x |e| i \bar \psi \left(\alpha +\beta \gamma^5 \right) \gamma^\mu \left( \partial_\mu + \frac{i}{2} \omega^{ab}_{\hphantom{ab}\mu} \gamma_{ab} \right) \psi + c.c. \, ,
\end{split}
\end{align}
where $|e|=\sqrt{|\gamma|}$ is the determinant of the vielbein fields, $\gamma_{ab}=[\gamma_a,\gamma_b]$  generate  Lorentz transformations in the spinor representation, $\{\alpha ,\beta\}$ are arbitrary coefficients and $\omega^{ab}_{\hphantom{ab} \mu}$ is the spin connection. The connection is minimally coupled to the fermions hence sources the spin current operator $i \bar \psi  \left(\alpha + \beta \gamma^5 \right) \gamma^\mu \gamma_{ab} \psi + cc.$.

When the sources in \eqref{diracLagrangian} are only background fields hence not subject to variations, there is no reason to choose the connection to be  Levi-Civita \footnote{Choosing the Levi-Civita connection is equivalent to symmetrizing the energy momentum tensor and removing the spin current via a pseudo-gauge transformation}. It can be more general 
\begin{equation}\label{LCK} 
\Gamma^\lambda_{\mu\nu} = \frac12 g^{\lambda\rho}\left(\partial_\mu g_{\nu\rho} + \partial_\nu g_{\mu\rho} - \partial_\rho g_{\mu\nu} \right) + K^\lambda_{\mu\nu}\, ,
\end{equation}
where the first term is the Levi-Civita connection and the second term, $K$ is called {\em contorsion}. The metricity requirement (vanishing covariant derivative of the metric) does not fix $K$. Instead, it is determined completely in terms of the {\em torsion} two-form $T$, which in turn is given by the covariant exterior derivative of the vielbein 
\begin{equation}\label{TCD} 
D e^a \equiv d e^a + \omega^a_b \wedge e^b  = T^a\, .
\end{equation}
In the presence of a non-trivial $T^a$, the vielbein and the spin connection stay independent. 

We can now extend the discussion from free fermions to a more general QFT with an action $I[e,\omega,\Psi]$ with $\Psi$ a non-trivial representation of the Lorentz algebra, coupled to a first order gravitational background. One then defines the effective action $W[e,\omega]$ 
\begin{align}\label{W}
\begin{split}
e^{i W[e,\omega]} = \int D \Psi e^{i I[e,\omega,\Psi]}\, ,
\end{split}
\end{align}
variation of which assumes the form 
\begin{align}
	\delta W = \int \left[\tau_a \delta e^a + \frac{1}{2} \sigma_{ab} \delta \omega^{ab} \right] \, .
\end{align}
We read the one point functions as 
\begin{align}\label{1pf1}
\begin{split}
\tau_a = \frac{\delta W[e,\omega]}{\delta e^a}, \quad \quad \sigma_{a b} = \frac{\delta W[e,\omega]}{\delta \omega^{ab}},
\end{split}
\end{align}
where the energy momentum three form $\tau_a$  and  the spin current three form $\sigma_{ab}$ are related to the standards currents by 
\vspace{1mm}
\begin{equation}\label{ts3}
T^{\mu \nu} =  \frac{\epsilon^{\nu \rho \sigma \lambda}}{|e|} e^{\mu b} \tau_{b, \rho \sigma \lambda}, \quad \quad S^\lambda_{\hphantom{\lambda} \mu \nu} = \frac{\epsilon^{\lambda \rho \sigma \tau}}{|e|} e^a_\mu e^b_\nu  \sigma_{a b, \rho \sigma \tau}.
\end{equation}
The Ward identities for the one-point functions (\ref{1pf1}) follow from invariance of $W$ under local Lorentz transformations and diffeomorphism transformations. Under local Lorentz
\begin{align}\label{localLorSym}
\begin{split}
		\delta_\lambda e^a &= - \lambda^a_{\hphantom{a}b} e^b \, , \qquad \qquad 
	\delta_\lambda \omega^{ab}= D\lambda^{ab}\, ,
\end{split}
\end{align}
where $D$ denotes the exterior covariant derivative on the tangent space. Invariance of $W$ results in
\begin{align}\label{lor1Eq}
	D \sigma^{ab} - 2 e^{[a} \tau^{b]} &= 0 \, .
\end{align}
Similarly, invariance under local diffeomorphism leads to 
\begin{align}
	\begin{split}
		\delta_\xi e^a &= \mathcal{L}_\xi e^a \, , \qquad \qquad 
		\delta_\xi \omega^{ab} = \mathcal{L}_\xi \omega^{ab} \, \, 
	\end{split}
\end{align}
the following equation of motion follows 
\begin{align}\label{diff1Eq}
	 D\tau_a - \left(I_a T^b \tau_b + \frac{1}{2}I_a R^{bc} \sigma_{bc} \right) + \frac{1}{2} I_a \omega^{cd} \left(D \sigma_{cd} - 2 e_{[c}\tau_{d]} \right) &= 0 \, .
\end{align}
with $I_a$ the contraction operator mapping p-forms into p-1 forms defined by $$I_a p =I_a \left(\frac{1}{p!} p_{a_1 ... a_n} e^{a_1}... e^{a_n} \right) =  \frac{1}{(p-1)!} p_{a ... a_n} e^{a_2}... e^{a_n} \, .$$  Equations \eqref{lor1Eq} and \eqref{diff1Eq} are the same as the conservation equations \eqref{hyd1} which also correspond to the relevant hydrodynamic equations in the next section. 

\vspace{2mm}
It is the object \eqref{W} that we want to compute using the holographic principle. Holography maps the effective action $W$ of a given large-N gauge theory in the strongly coupling regime to the semi-classical gravitational action $I[e,\omega]_{\text{on-shell}}$ that is evaluated on the corresponding gravitational background and renormalized using holographic renormalization \cite{Skenderis,deHaro:2000vlm}: 
\begin{equation}\label{fieldOnShell}
 I_{\text{on-shell}} = W
\end{equation}
We assume, as in \cite{Banados}, that the holographic principle holds for asymptotically AdS solutions in the LCS gravity and compute the conserved currents by varying the on-shell gravitational action. Finally, the Ward identities  \eqref{lor1Eq} and \eqref{diff1Eq} in the holographic dual follow from invariance of the gravitational action $I_{\text{on-shell}}$ under asymptotic symmetries, as we review in section \ref{sec::Noether}. 

\section{Hydrodynamics with spin current}\label{sec::hydro}

Hydrodynamics is the long-wavelength, large distance effective theory of (\ref{W}). In this limit we expect  the currents $T^{\mu \nu}$ and $S^\lambda_{\hphantom{\lambda} \mu \nu}$ to be the only relevant quantities and the corresponding hydrodynamic equations are given by the Ward identities \eqref{lor1Eq} and \eqref{diff1Eq} which we write in the coordinates of the space-time where the fluid is embedded as 
\begin{align}\label{hyd2}
\begin{split}
\nabla_\mu(|e| T^{\mu}_{\nu}) = \frac{|e|}{2} S^{\lambda}_{\hphantom{\lambda} \rho \sigma} R^{\rho \sigma}_{\hphantom{\mu \nu} \lambda \nu } \, , \quad \quad \quad
\nabla_\lambda \left(|e| S^\lambda_{\hphantom{\lambda} \rho \sigma}  \right) = \frac{|e|}{2} T_{[\rho \sigma]} \, ,
\end{split}
\end{align}
The energy-momentum tensor is not conserved because of the external force that arise from the curvature. This force term is completely analogous to the electromagnetic force $j^\mu F^{\mu\nu}$ in the presence of an external electromagnetic field; S being analogous to j and R to F. 

The fluid is described by the four velocity vector $u^\mu$, normalized as $u^\mu u_\mu = -1$, evolving on a background specified by a metric $\gamma_{\mu \nu} = \theta^a_\mu \theta^b_\nu \eta_{ab}$ and an independent connection $\omega^{ab}_{\hphantom{ab}\mu}$. Any tensor can be decomposed into its projection along and tangent to $u^\mu$ using the projector
\begin{align}
\Delta_{\mu \nu}&=u_\mu u_\nu + \gamma_{\mu \nu} \, ,
\end{align}
which satisfies $\Delta_{\mu \nu} u^\mu = 0$ and $\Delta_{\mu \nu} \Delta^\nu_\rho = \Delta_{\mu \rho}$. Our goal in this section is to perform this decomposition for the currents $\{T^{\mu \nu}, S^\lambda_{\hphantom{\lambda}\mu \nu} \}$. For simplicity we will consider a flat background in the metric sense\footnote{Note that we define the derivative counting with respect to $\partial$ only, instead of the $\nabla$ which also involves torsion. This means the independent connection $\omega^{ab}_\mu$ is taken to be $\mathcal{O}\left(1\right)$ in the derivative counting.}, namely we will take $\gamma_{\mu \nu}=\eta_{\mu \nu}$. In this low energy limit the currents can then be organized in a gradient expansion as 
\begin{align}
	\begin{split}
		T^{\mu \nu} &= T^{(0) \mu \nu} + T^{(1) \mu \nu} + ... \\
		S^{\lambda \mu \nu} &= S^{(0) \mu \nu} + S^{(1) \lambda \mu \nu} + ...
	\end{split}
\end{align}
where $(n)$ indicates $\mathcal{O}\left( \partial^n \right)$ in the derivative expansion. We then introduce a generic constitutive relations for these currents
\begin{align}
	\begin{split}\label{currentsVar}
		T^{(n) \mu \nu}&= \sum_l^{l_n} b_{l} \, m^{\mu \nu}_{l} \, , \\
		S^{(n) \lambda \mu \nu} &= \sum_k^{k_n} c_k\,  m^{\lambda \mu \nu}_k  \, , 
	\end{split}
\end{align}
with $ \{m^{\mu \nu}_l, m_k^{\lambda \mu \nu} \} $ representing the $l_n$ independent rank 2 and $k_n$ independent rank 3 tensors containing $m$ derivatives which are constructed from the sources $\{e^a,\omega^{ab}\}$. The  coefficients $\{b_l, c_k \}$ will be the $l_n+ k_n$ transport coefficients at $\mathcal{O}\left( \partial^m \right)$. A detail classification of transport coefficients for non trivial spin sources, including possible relations among them\footnote{These relations typically arise from Onsager relations or positivity of the divergence of the entropy current.} is unknown to us and beyond the scope of this work. Instead, we explicitly compute the currents in  \eqref{currentsVar} up to $\mathcal{O}\left( \partial \right)$ and identify the transport coefficients from them. 

\subsection{Hydrodynamic decomposition of spin sources} \label{sec::IrreSpin}

The external source for the spin current in the effective action (\ref{W}) is given by the contorsion $\omega^{ab}$ which is the only non-vanishing part of the spin connection on a flat background. To better understand its content it is useful to decompose in terms of irreducible representations of the boundary Lorentz algebra \cite{Torsion}. This decomposition for $\omega^{ab}$ yields three irreducible representations of SO(3,1): a vector $\left(\frac{1}{2},\frac{1}{2}\right)$, an axial vector $\left(\frac{1}{2},\frac{1}{2}\right)$ and a tensor $\left(\frac{3}{2},\frac{1}{2}\right)\oplus\left(\frac{1}{2},\frac{3}{2}\right)$:
\begin{align}
\begin{split}\label{preDecomposeK}
 \omega^{ab} = \omega^{ab}_{\text{vector}} + \omega^{ab}_{\text{axial}} + \omega^{ab}_{\text{tensor}}  \, .  
\end{split}
\end{align}
We can now further decompose (\ref{preDecomposeK}) with respect to $u^\mu$. Going to the rest frame of the fluid $u^\mu = (1,\vec{0})$, one sees that this is equivalent to decomposing according to rotation subgroup of the local Lorentz symmetry: 
\begin{itemize}
\item Vector: $\left( \frac{1}{2}, \frac{1}{2} \right) = 1 \oplus 0$ - A scalar and a vector (4 independent degrees of freedom)
\item Axial:  $\left( \frac{1}{2}, \frac{1}{2} \right) = 1 \oplus 0$ - A pseudo-scalar and a pseudo-vector (4 independent degrees of freedom)
\item Tensor: $\left(\frac{3}{2},\frac{1}{2} \right) \oplus \left(\frac{1}{2},\frac{3}{2} \right)=1 \oplus 2 \oplus 1 \oplus 2$ - A vector, a pseudo-vector, a symmetric traceless rank 2 tensor and a symmetric traceless rank 2 pseudo tensor (16 independent degrees of freedom).  
\end{itemize}
This decomposition can be made explicit for the vector and axial parts of the contorsion by parametrizing it through the vector field $\tilde V$  and the axial field  $\tilde  A$ as
\begin{align}
\begin{split}\label{cont1}
\omega^{ab}_{\text{vector}} &= \theta^a_\mu \theta^b_\nu \left[ \delta^\mu_\beta \tilde V^\nu - \delta^\nu_\beta \tilde V^\mu    \right] dx^\beta \, ,
\end{split}
\end{align}
\begin{align}
\begin{split}\label{cont2}
\omega^{ab}_{\text{axial}} &= \theta^a_\mu \theta^b_\nu \left[ \epsilon^{\mu \nu \rho \sigma} \tilde A_\rho \gamma_{\sigma \beta}    \right] dx^\beta \, .
\end{split}
\end{align}
The vector and axial fields $\tilde V$ and $\tilde A$ can be further decomposed with respect to the four velocity $u^\mu$ as 
\begin{align}\label{projections0}
\tilde V = - \mu_V  u + V, \quad \quad \quad  \, ,
\tilde A = - \mu_A u + A \, , 
\end{align}
where $\mu_V$ and $\mu_A$ can be thought of vector and axial chemical potentials, and the vectors $V$ and $A$ are orthogonal to the four velocity,
\begin{align}
A^\mu u_\mu =0 \, , \quad \quad V^\mu u_\mu =0 \, .
\end{align}
The remaining irreducible tensor corresponds to the traceless and pseudo-traceless part of the connection and in general can be written as
\begin{align}
\begin{split}\label{cont3}
\omega^{ab}_{\text{tensor}}\equiv \theta^a_\mu \theta^b_\nu\, \omega^{\mu \nu}_{T \gamma} dx^\gamma= \omega^{ab}-\omega^{ab}_{\text{vector}} -\omega^{ab}_{\text{axial}}\, .
\end{split}
\end{align}
The general hydrodynamic decomposition of $\omega^{\mu \nu}_{T \beta}$ are given by
\begin{align}
\begin{split}\label{projections}
\omega^{\mu \nu}_{T \gamma} \Delta_{\alpha \mu} \Delta_{\beta \nu} \Delta^\gamma_\lambda &= -\epsilon_{\mu \alpha \beta \rho} u^\mu   \left[ \mathcal{C}^{\rho \sigma} - \frac{1}{2} \epsilon^{\nu \rho \sigma \tau} u_\nu  D_\tau  \right] \Delta_{\sigma \lambda}\, , \\
\omega^{\mu \nu}_{T \gamma} u_\mu \Delta_{\alpha \nu} \Delta^{\gamma}_\beta &= \mathcal{H}_{\alpha \beta} - \frac{1}{2} \epsilon_{\mu \alpha \beta \tau} u^\mu W^\tau \, , \\
\omega^{\mu \nu}_{T \gamma} \Delta_{\mu \alpha} \Delta_{\nu \beta} u^\gamma &= \epsilon_{\mu \alpha \beta \tau} u^\mu W^\tau \, , \\
\omega^{\mu \nu}_{T \gamma} u_\mu \Delta_{\nu \alpha} u^\gamma &= D_\alpha \, , 
\end{split}
\end{align}
where the traceless and pseudo-traceless conditions have been implemented, the tensors $\mathcal{C}_{\rho \sigma}$ and $\mathcal{H}_{\alpha \beta}$ are symmetric and traceless, $D$ is a vector, $W$ is an axial vector,  and the indices of all the tensor fields $\{ \mathcal{C}_{\alpha \beta},\mathcal{H}_{\alpha \beta},D_\alpha, W_\alpha \}$ are orthogonal to the four velocity $u$,
\begin{align}
\begin{split}
\mathcal{C}_{\alpha \beta} u^\alpha = 0\, ,\quad \quad \quad \mathcal{H}_{\alpha \beta} u^\alpha =0\, , \quad \quad \quad
D_\alpha u^\alpha =0\, , \quad \quad \quad   W_\alpha u^\alpha =0\, ,
\end{split}
\end{align}
all together the tensor fields  $\{ \mathcal{C}_{\alpha \beta},\mathcal{H}_{\alpha \beta},D_\alpha, W_\alpha \}$ represent 16 independent components which agrees with the expected independent components of the tensor irreducible representation, see the table at the end of this section. We call this decomposition the ``irreducible decomposition" in what follows. This decomposition \eqref{projections} of the sources is the most general one as all the independent degrees of freedom are taken into account and classified according to the rotational symmetry of the fluid rest frame: a scalar  $\mu_V$, a pseudo-scalar $\mu_A$, two vectors $\{V^\mu, D^\mu\}$, two pseudo-vectors $\{A^\mu, W^\mu\}$, a symmetric rank 2 tensor $\mathcal{H}^{\mu \nu}$ and a symmetric rank 2 pseudo-tensor $\mathcal{C}^{\mu \nu}$. 

An alternative regrouping of the irreducible components --- which we call the ``hydrodynamic decomposition" is obtained  by decomposing with respect to the four velocity and its projector as
\begin{align}
\begin{split}\label{decompConnection}
\omega^{ab}&= \theta^a_\mu \theta^b_\nu \left[ -\epsilon^{\mu \nu}_{\hphantom{\mu \nu} \rho \sigma} u^\rho \left( \mu_A \delta^{\sigma}_{\beta} + \mathcal{C}^{\sigma}_{\beta}- \mathcal{A}_{1}^\sigma u_\beta \right) + 2 \left( \mu_V u^{[\mu} -  \mathcal{V}^{[\mu}_2 \right) \Delta^{\nu]}_\beta \right. \\
&\hphantom{=} \hphantom{\theta^a_\mu \theta^b_\nu [[[} \left. +  2 u^{[\mu}\epsilon^{ \nu]}_{\hphantom{\nu]} \rho \sigma \beta} u^\rho \mathcal{A}^\sigma_2 - 2 u^{[\mu}\left(\mathcal{V}^{\nu]}_1 u_{\beta} + \mathcal{H}^{\nu]}_\beta  \right) \right] dx^\beta \, ,
\end{split}
\end{align}
with $\{ \mathcal{A}_1, \mathcal{A}_2, \mathcal{V}_1, \mathcal{V}_2 \}$ defined as 
\begin{align}
\begin{split}
\mathcal{A}^\mu_1 &= A^\mu - W^\mu  \, , \\
\mathcal{A}^\mu_2  &= A^\mu + \frac{W^\mu}{2} \, , \\
\mathcal{V}^\mu_1  &= V^\mu - D^\mu  \, , \\
 \mathcal{V}^\mu_2  &= V^\mu + \frac{D^\mu}{2}\, .
\end{split}
\end{align}
We refer to the sources $\{V,A,D,W \}$ as the irreducible vector/axial sources and to the sources $\{\mathcal{V}_1,\mathcal{V}_2,\mathcal{A}_1,\mathcal{A}_2 \}$ as the hydrodynamic vector/axial sources. As shown in section \ref{sec::ZerothSolution} the hydrodynamic sources appear naturally when solving the constraint equations. 

\subsection{Hydrodynamic decomposition of the energy-momentum and spin currents}\label{sec::SpinHydro}

We consider a 3+1 dimensional background and start with decomposing the energy momentum tensor. As clear from (\ref{hyd2}) we should allow for an antisymmetric part of the energy-momentum tensor: 
\begin{align}\label{compTmn}
T^{\mu \nu} = \varepsilon u^\mu u^\nu - p \Delta^{\mu \nu} + \bar q^\mu u^\nu + u^\mu q^\nu + \pi^{\mu \nu} + \tau^{\mu \nu} \, ,
\end{align}
with $\varepsilon$ the energy density, $p$ the pressure, $q^\mu$ and $\bar q^\mu$ energy currents, $\pi^{\mu \nu}$ the symmetric traceless tensor including shear, and $\tau^{\mu \nu}$ the antisymmetric part which we refer to as the intrinsic torque. They are obtained in terms of the projections 
\begin{align}
\begin{split}\label{projectionsT}
\varepsilon &=u_\mu u_\nu T^{\mu \nu} \, ,\\
p &=- \frac{1}{3} \Delta_{\mu \nu} T^{\mu \nu} \, , \\
\bar q^\mu &= -\Delta^\mu_\rho u_\sigma T^{\rho \sigma} \, , \\
q^\nu &=-u_\rho \Delta^\nu_\sigma T^{\rho \sigma} \, , \\
\pi^{\mu \nu} &= \left( \Delta^\mu_{(\rho}\Delta^\nu_{\sigma)} - \frac{1}{3} \Delta^{\mu \nu} \Delta_{\rho \sigma}  \right) T^{\rho \sigma} \, , \\
\tau^{\mu \nu} &= \Delta^\mu_{[\rho} \Delta^\nu_{\sigma]}  T^{\rho \sigma}\, .
\end{split}
\end{align}
We note that $\{\bar q^\mu, q^\nu, \pi^{\mu \nu},\tau^{\mu \nu} \}$ are orthogonal to the four velocity. Whenever $\Delta q^\mu \equiv \bar q^\mu - q^\mu$ and $\tau^{\mu \nu}$ are non vanishing the spin current will not be conserved, see (\ref{hyd2}). In a 3+1 dimensional spacetime the spin current itself can decomposed as follows
\begin{align}
\begin{split}\label{spinDecomp}
S^{\lambda \mu \nu} &=   u^\lambda \left(  u^\nu n_V^\mu -u^\mu n_V^\nu  + \epsilon^{\mu \nu}_{\hphantom{\mu \nu} \rho \sigma}u^\rho n^{\sigma}_{A}   \right)  + \rho_A \epsilon^{\lambda \mu \nu \rho} u_\rho +\rho_V \left( \Delta^{\lambda \nu} u^\mu - \Delta^{\lambda \mu} u^{\nu} \right) \\ &\hphantom{=} +  N^{\lambda \mu} u^{\nu}-N^{\lambda \nu}u^\mu - \epsilon^{\mu \nu \rho \sigma} u_\rho \bar N^\lambda_\sigma +\Delta^{\lambda \nu} \bar n^\mu_V   - \Delta^{\lambda \mu}\bar n_V^{\nu}   \\ &\hphantom{=} +  \epsilon^{\lambda \mu}_{\hphantom{\lambda \mu} \rho \sigma} u^{\nu} u^\rho \bar n_{A}^\sigma - \epsilon^{\lambda \nu}_{\hphantom{\lambda \mu} \rho \sigma} u^\mu u^\rho \bar n_A^\sigma \, ,
\end{split}
\end{align}
with $\{\rho_V,\rho_V\}$ vector and axial charge densities, $\{n^\mu_V, \bar n^\mu_V \}$  and  $\{n^\mu_A, \bar n^\mu_A \}$ the vector and axial currents orthogonal to the four velocity,  and $\{N^{\mu \nu}, \bar N^{\mu \nu}  \}$  symmetric traceless tensor and pseudo-tensor currents orthogonal to the four velocity. These currents and densities can be obtained from the projections
\begin{align}
\begin{split}
\rho_V &= \frac{1}{3} \Delta_{\rho \alpha} u_\beta  S^{\rho \alpha \beta} \, ,\\
\rho_A &= \frac{1}{3!} \epsilon_{\alpha \beta \rho  \sigma} u^\sigma S^{\rho \alpha \beta} \, , \\
n^\mu_V &= u_\rho u_\beta \Delta^\mu_\alpha S^{\rho \alpha \beta} \, ,\\
\bar n^\mu_V &= \frac{1}{2} \Delta^\mu_\beta \Delta_{\rho \alpha} S^{\rho \alpha \beta} \, \\
n^\mu_A &= \frac{1}{2} \epsilon^\mu_{\hphantom{\mu} \alpha \beta \sigma}u^\sigma u_\rho S^{\rho \alpha \beta} \, ,\\
\bar n^\mu_A &= \frac{1}{2}\epsilon^\mu_{\hphantom{\mu} \beta \rho \sigma} u^\sigma u_\alpha S^{\rho \alpha \beta} \, , \\
N^\lambda_\kappa &= u_\alpha \left(\Delta^{\lambda}_{(\beta}\Delta_{\rho) \kappa} - \frac{1}{3} \Delta^\lambda_\kappa \Delta_{\beta \rho} \right) S^{\rho \alpha \beta}\, ,\\
\bar N^\lambda_\kappa &= \frac{1}{3!} u^\sigma \left(   3 \epsilon^\lambda_{\hphantom{\lambda} \kappa \beta \sigma } \Delta_{\rho \alpha} + 3 \Delta^\lambda_\rho \epsilon_{\kappa \alpha \beta \sigma}-\Delta^\lambda_\kappa \epsilon_{\rho \alpha \beta \sigma} \right) S^{\rho \alpha \beta} \, , \\
\end{split}
\end{align}
Decomposition \eqref{spinDecomp} is analogous to the decomposition \eqref{decompConnection} of the spin connection which we call the hydrodynamic decomposition. 

There exist an equivalent ``irreducible" decomposition of the spin current in terms of its irreducible components under rotation, analogous to the decomposition for the spin connection \eqref{preDecomposeK}. This consists of an axial current $J^\mu_A$, a vector current $J^\mu_V$, and a tensor spin current which further can be decomposed into a tensor current $N^{\mu \nu}$, a pseudo-tensor current $\bar N^{\mu \nu}$, a vector current $\bar J^\mu_V$ and an axial vector current $\bar J^\mu_A$. The tensor currents are the same in both the hydrodynamic and irreducible decompositions while the vector/axial currents from the irreducible decomposition are related to the vector/axial currents and densities in the hydrodynamic decomposition by
\begin{align}
\begin{split}\label{irreCurrents}
J^\mu_V &=-\rho_V u^\mu+\frac{1}{3} \left( n^\mu_V + 2 \bar n^\mu_V \right) \, ,\\
\bar J^\mu_V &= \frac{2}{3} \left(\bar n^\mu_V - n^\mu_V \right) \, ,\\
J^\mu_A &= -\rho_A u^\mu+\frac{1}{3} \left( n^\mu_A + 2 \bar n^\mu_A \right) \, , \\
\bar J^\mu_A &= \frac{2}{3} \left(\bar n^\mu_A - n^\mu_A \right) \, .
\end{split}
\end{align}
From the currents of \eqref{irreCurrents} is of particular interest the axial one $J^\mu_A$. When the canonical fields are Dirac fermions it corresponds to the expectation value of the standard axial current operator $J^\mu_A = \langle \bar \psi \gamma^\mu \gamma_5 \psi \rangle$. We  summarize the coupling between spin sources and the hydrodynamic components of the spin current in the following table:

\begin{table}
	\begin{tabular}{| l | c | r| }
  \hline			
  Spin Source & Spin current & Degrees of Freedom \\ \hline
  $\mu_V$ & $\rho_V$ & 1 \\ \hline
  $\mu_A$ & $\rho_A$ & 1 \\ \hline 
  $\mathcal{V}^\mu_1$ & $n^\mu_V$ & 3 \\ \hline 
  $\mathcal{A}^\mu_1$ & $n^\mu_A$ & 3 \\ \hline 
  $\mathcal{V}^\mu_2$ & $\bar n^\mu_V$ & 3 \\ \hline 
  $\mathcal{A}^\mu_2$ & $\bar n^\mu_A$ & 3 \\ \hline 
  $\mathcal{H}^{\mu \nu}$ & $N^{\mu \nu}$ & 5 \\ \hline 
  $\mathcal{C}^{\mu \nu}$ & $\bar N^{\mu \nu}$ & 5 \\ \hline 
\end{tabular}
\caption{Hydrodynamic decomposition of the spin sources and the spin current. Each source is paired with the component of the spin current it sources together with their corresponding number of degrees of freedom.} \label{table}
\end{table}

\section{Holographic 5D Lovelock Chern-Simons gravity}\label{sec::Holography}

The holographic backgrounds that we consider in this work as dual to quantum field theories with a non-trivial spin current, are 5-dimensional backgrounds that solve the Chern-Simons action
\begin{align}\label{CSActionForm}
S=  \int \langle \mathcal{F} \wedge \mathcal{F} \wedge \mathcal{A} - \frac{1}{2} \mathcal{F}\wedge \mathcal{A}\wedge \mathcal{A}\wedge \mathcal{A} + \frac{1}{10} \mathcal{A}\wedge\mathcal{A}\wedge\mathcal{A}\wedge\mathcal{A}\wedge\mathcal{A}\rangle \, ,    
\end{align}
where  $\langle \rangle$  indicates a group trace over the SO(4,2) algebra with generators $\mathcal{J}_{\bar A}$ on which the connection one form $\mathcal{A}$ is valued, and, where the field strength is defined as usual $\mathcal{F}=d \mathcal{A} + \mathcal{A} \wedge \mathcal{A}$. It was shown in \cite{Banados} that, for this five dimensional gravitational Chern-Simons (CS) theory\footnote{In this paper finite Fefferman-Graham expansion was also established for the 3D CS theory.} a certain class of gauge allows for a finite Fefferman-Graham (FG) expansion. Therefore a well defined holographic recipe can be established for this theory. Their work was generalized in \cite{Miskovic} to any odd dimension gravitational CS theory where an analysis of the boundary gauge symmetries was also performed. 

Here, we first outline the necessary details of the theory. Our discussion closely follows the presentation in \cite{Banados}. The SO(4,2) indices $\bar A$ can be written as a pair of antisymmetric indices via $\bar A = \{ AB, A6 \}$ with the indices $A=\{0,1,2,3,5\}$. Using this index decomposition the following identification can be done $\{ \mathcal{J}_{A 6}= P_A , \mathcal{J}_{AB}=J_{AB}\}$ with $P_A$ and $J_{AB}$ the generators of translations and Lorentz transformations in five dimensions satisfying the SO(4,2) algebra 
\begin{align}
\begin{split}
\left[P_A,P_B \right] &= J_{AB}\, , \\
\left[ P_A, J_{BC} \right] &= \eta_{AB} P_C - \eta_{AC} P_B \, , \\
\left[ J_{AB}, J_{CE} \right] &= \eta_{BC} J_{AE} + \eta_{AE} J_{BC} - \eta_{AC} J_{BE} - \eta_{BE} J_{AC}\, .
\end{split}
\end{align}
The action \eqref{CSActionForm} is invariant under infinitesimal gauge transformations
\begin{align}
\begin{split}\label{infiGauge}
\delta \mathcal{A} = d \tau + \left[ \mathcal{A},\tau \right]\, , \\
\tau= \eta^A P_A + \frac{1}{2} \lambda^{AB} J_{AB}\, ,
\end{split}
\end{align}
with gauge parameters $\{\eta^A, \lambda^{AB} \}$ representing local translations and Lorentz transformations,  up to a boundary term
\begin{align}
\delta S = -\frac{1}{2} \int_{\partial \mathcal{M}_5} \langle d \tau \wedge \left(\mathcal{A}\wedge \mathcal{A} + d\mathcal{A} \wedge\mathcal{A} + \mathcal{A}\wedge \mathcal{A} \wedge\mathcal{A} \right) \rangle \, .
\end{align}
\vspace{2mm}
To connect to the five dimensional gravity the 5D vielbein\footnote{Below we denote the 5D quantities with a hat symbol to distinguish them from the corresponding quantities in 4D.} $\hat e^A$ and the spin connection $\hat \omega^{AB}$  are identified as components of the gauge connection $\mathcal{A}$ via \cite{Chamseddine} 
\begin{align}
\begin{split}\label{gaugeID}
\mathcal{A} &\equiv \hat e^A P_A + \frac{1}{2} \hat \omega^{AB} J_{AB} \, , \\
\mathcal{F} &= \hat T^A P_A + \frac{1}{2} \left(R^{AB} + \hat e^A \hat e^B \right) J_{AB} \, ,
\end{split}
\end{align}
where $\hat T^A$ and $\hat R^{AB}$ are the five dimensional torsion and curvature two forms defined as
\begin{align}
\begin{split}
\hat T^A &= d \hat e^A + \hat \omega^A_{\hphantom{A}B} \wedge \hat e^B\, , \\
\hat R^{AB} &= d \hat \omega^{AB} + \hat \omega^A_{\hphantom{A}C}  \wedge \hat\omega^{CB}\, .
\end{split}
\end{align}
Using the identification \eqref{gaugeID} the CS action \eqref{CSActionForm} can be written in the more familiar form
\vspace{2mm}
\begin{align}\label{SO42}
\begin{split}
S_{LCS} &= \kappa \int_{M_5} \epsilon_{ABCDE} \left[ \hat R^{AB} \hat  R^{CD}  \hat e^E  + \frac{2}{3} \hat R^{AB}  \hat e^C  \hat e^D   \hat e^E   + \frac{1}{5} \hat e^A \hat e^B \hat e^C \hat e^D \hat e^E \right]\, ,
\end{split}
\end{align}
where for notational simplicity we omit the wedge product symbol. Here $\kappa$ is the CS parameter that arises from the non-vanishing group trace
\begin{align}
\begin{split}\label{symmetricTrace}
\langle  \mathcal{J}_{ A  B} \mathcal{J}_{ C  D} \mathcal{J}_{E 6} \rangle = \frac{\kappa}{2} \epsilon_{ A  B  C  D  E  } \, .
\end{split}
\end{align}
The equations of motion for action \eqref{SO42} are 
\begin{align}\label{eqm1}
\epsilon_{ABCDE}\left(\hat R^{AB} + \hat e^A  \hat e^B \right) \left(\hat R^{AB} + \hat e^A  \hat e^B  \right)&=0 \, , \\ \label{eqm2}
\epsilon_{ABCDE}\left( \hat R^{AB} + \hat e^A  \hat e^B  \right) \hat T^E&=0 \, ,
\end{align}
and can be compactly written in the CS form 
\begin{equation}\label{CSEquation}
g_{\bar A \bar B \bar C} \mathcal{F}^{\bar B} \mathcal{F}^{\bar C}=0 \, , 
\end{equation}
where $g_{\bar A \bar B \bar C}= \langle \mathcal{J}_{\bar A} \mathcal{J}_{\bar B} \mathcal{J}_{\bar C} \rangle$ is the trilinear symmetric invariant trace whose only non-vanishing component is given by \eqref{symmetricTrace}. There exist non-trivial solutions to \eqref{eqm1} and \eqref{eqm2} with non-vanishing torsion $\hat T^A\neq 0$, see \cite{Garay1, Garay2}. We will indeed find such novel solutions in the section \ref{sec::ZerothSolution} and \ref{sec::FirstSolution}.

\subsection{Gauge Fixing and Fefferman-Graham expansion} 

We look for asymptotically AdS solutions to  \eqref{eqm1} and \eqref{eqm2} which to find holographic duals to fluids with spin degrees of freedom. We work with a radial foliation of the spacetime coordinates $x^M=(x^\mu,r )$ and similarly of the tangent space indices $A= (a,5)$ where the asymptotic boundary of the manifold is located at $r=r_0$ which we send to infinity $r_0 \rightarrow \infty $ after holographic renormalization.  For metric formulations of gravity Fefferman-Graham (FG) theorem tell us that the metric near this boundary takes the form \cite{Fefferman} 
\begin{align}\label{Fefferman-Graham}
ds^2 &= \frac{d\rho^2}{4 \rho^2} + \frac{1}{\rho^2} g_{\mu \nu}(x^\mu,\rho) dx^\mu dx^\nu\, , 
\end{align}
where we have introduced the FG coordinate $\rho=\frac{1}{r^2}$. On even d-dimensional spacetimes $g_{\mu \nu}$ admits the expansion  
\begin{align}
g_{\mu \nu}(x^\mu, \rho) &= g^0_{ \mu \nu}(x^\mu) + ... + \rho^{\frac{d}{2}} g^d_{ \mu \nu}(x^\mu) \, .
\end{align}
To derive the equivalent of \eqref{Fefferman-Graham} in the first order formulation, it is convenient to use the CS form \eqref{CSEquation} of the equations of motion \cite{Banados}  and note that under the radial foliation they split as 
\begin{align}\label{cEquation}
g_{\bar A \bar B \bar C} \epsilon^{\mu \nu \alpha \beta} \mathcal{F}^{\bar B}_{\hphantom{\bar B} \mu \nu} \mathcal{F}^{\bar C}_{\hphantom{\bar C}\alpha \beta}&= 0 \, , \\
g_{\bar A \bar B \bar C} \epsilon^{\mu \nu \alpha \beta} \mathcal{F}^{\bar B}_{\hphantom{\bar B} \mu \nu} \mathcal{F}^{\bar C}_{\hphantom{\bar C}\alpha r}&= 0\, . \label{dEquation} 
\end{align}
Equation \eqref{cEquation} contains no radial derivative and can be regarded as a constraint that holds at every value of $r$. It was shown in \cite{Banados} that the Ward identities of the  dual CFT are contained in this set of constraints. The bulk dynamics of the theory are contained in \eqref{dEquation} and it is from this expression that the FG structure arises. Equations \eqref{cEquation} and \eqref{dEquation} are not independent and for a generic solution of \eqref{cEquation} it follows that \eqref{dEquation} implies \cite{CSConstraint}

\begin{align}\label{gaugeChoice1}
\mathcal{F}^{\bar B}_{\hphantom{\bar A} r \mu } = \mathcal{F}^{\bar B}_{\hphantom{\bar B} \mu \nu}N^\nu \, ,
\end{align}
with $N^\nu$ arbitrary functions. Equation \eqref{gaugeChoice1} can be rewritten as 
\begin{align}\label{dynamicalEq}
\partial_r \mathcal{A}^{\bar B}_{\hphantom{\bar B} \mu}= D_\mu \mathcal{A}^{\bar B}_{\hphantom{\bar B} r}+ \mathcal{F}^{\bar B}_{\hphantom{\bar B}\mu \nu} N^\nu \, , 
\end{align}
where $D_\mu$ denotes a covariant derivative. We note that the right hand side of \eqref{dynamicalEq} corresponds to a gauge transformation parametrized by $\mathcal{A}_r$ and a diffeomorphism\footnote{We are using the gauge invariant  form for diffeomorphisms \cite{Jackiw}  that differs from a Lie derivative by a local gauge transformation with parameter $\tau= \mathcal{A}_\mu \xi^\mu$.} with parameter $N^\nu$. This means $A_\mu(r+\delta r)$ is determined from $A_\mu(r)$ by means of a gauge transformation and a diffeomorphism, allowing us to choose the functions $\{ \mathcal{A}_r, N^\nu\}$  at will. By fixing diffeomorphisms on the transverse direction $x^\mu$ we can set $N^\mu =0 $ leaving only $\mathcal{A}_r$ to be fixed by a gauge transformation. 

One crucial remark here is that different gauge choices for $\mathcal{A}_r$ give rise to non-equivalent boundary theories \cite{Miskovic} as the CS action is only gauge invariant up to boundary terms. We consider the following parametrization of the gauge choice $\mathcal{A}_r$
\begin{align}
\begin{split}\label{gaugeChoiceGeneral}
\mathcal{A}_r &= H(r,x) P_5 + H^a_+(r,x) J_a^+ + H^a_-(r,x) J_a^- +\frac{1}{2} H^{ab}(r,x) J_{ab}\, ,
\end{split}
\end{align}
with $J^{\pm}_a \equiv P_a \pm J_{a5}$ and where $\{H,H^a_{\pm}, H^{ab}\}$ are arbitrary functions of the holographic coordinate and of the boundary coordinates. Setting $\{H^a_{\pm},H^{ab} \}$ to zero and $H=H(r)$ corresponds to the gauge choice\footnote{The simplest gauge sets $\{H,H^a,H^{ab}\}$ to zero, however this will result in a degenerate metric.} in \cite{Banados,Miskovic}. Here we will work with a slight generalization by allowing $H=H(r,x)$, more general gauge choices will be treated elsewhere \cite{GallegosNew}. In particular we assume $H(r,x)$ to assume the following form 
\begin{align}\label{HasG}
	H= \frac{1}{r g(r,x)} \ ,
\end{align}
with $g(r,x)$ playing the role of a blackening factor which in principle be fixed by making use of the remaining radial diffeomorphism. In particular, in \cite{Banados} it is set to 1 with the corresponding solution being global AdS. However when $g(r,x)$ has poles, as in a blackhole, the radial diffeomorphism required to set $g$ to unity would be singular.  These give rise to a class of blackhole solutions distinct from the global AdS solutions considered in \cite{Banados,Miskovic}. We therefore consider $g(r,x)$ with simple poles. A generic near boundary asymptotic is then given by
\begin{align}\label{expansionBlackening}
	g(r,x) = 1 + \sum \frac{c_i(x)}{r^{i+1}}  \, ,
\end{align}
with $c_i(x)$ some real functions. The solution to \eqref{dynamicalEq} in this gauge becomes
\begin{equation}\label{solutionCS}
\mathcal{A}(\rho,x) = e^{-P_5 \int H(r,x) d\rho } \mathcal{A}(0,x) e^{P_5 \int H(r,x) d\rho} + P_5 \left[dx^\mu \partial_\mu  \int H(r,x) \right] \, .
\end{equation}
For convenience we switched back to the FG coordinates and where $\mathcal{A}(0,x)$ is the boundary condition for the gauge connection at the AdS boundary. One finds, using \eqref{HasG} and \eqref{expansionBlackening}, that the function $H$ when integrated over the radial coordinate satisfies the following asymptotic expansion
\begin{align}
\begin{split}\label{HExpansion}
\int H d \rho &= -\frac{\ln \rho}{2} + c_0 \rho^{1/2}   + \frac{c_1-c^2_0 }{2}\rho  + \frac{c_0^3-2c_0 c_1 + c_2}{3} \rho^{3/2}  \\ & \hphantom{=} +  \frac{c_3 - c^4_0 + 3 c^2_0 c_1 - c^2_1 - 2 c_0 c_2}{4} \rho^2 +... \, .
\end{split}
\end{align}
On the other hand, using the gauge choice discussed above, the boundary condition  $A(0,x)$ can be parametrized as 
\begin{equation}\label{initialCond}
A(0,x) = e^a(x) J^+_a + k^a(x) J^-_a + \frac{1}{2} \omega^{ab}(x) J_{ab} \, ,
\end{equation}
Combining \eqref{solutionCS}, \eqref{initialCond} and \eqref{HExpansion} the near boundary expansion of $\hat e^{M}$ and $\hat \omega^{MN}$ is 
\begin{align}\label{asympSol}
\begin{split}
\hat e^5= -\frac{d \rho}{2 \rho}\, , \quad \quad  \hat  e^a = \frac{1}{\sqrt{\rho}} \left[ \tilde e^a +   \rho \tilde k^a  \right]\, , \quad \quad  
\hat  \omega^{a5} = \frac{1}{\sqrt{\rho}} \left[ \tilde e^a  - \rho \tilde k^a  \right]\, , \quad \quad  
\hat  \omega^{ab}=\omega^{ab} \, ,
\end{split}
\end{align}
with $\tilde e^a$  and $\tilde k^a$ a shorthand notation for 
\begin{align}
\begin{split}
\tilde e^a &= \left[1 + c_0 \rho^{1/2} + \frac{c_1}{2} \rho + \frac{2 c_2 - c_0 c_1}{6} \rho^{3/2} + \frac{6 c_3 - 4 c_0 c_2 - 3 c_1^2 + 2 c_0^2 c_1}{24} \rho^2 + ... \right] e^a \, , \\
\tilde k^a &= \left[1 - c_0 \rho^{1/2} + \frac{2c^2_0- c_1}{2}\rho + ... \right]k^a \, .
\end{split}
\end{align}
Here we only show the terms that are relevant for the thermodynamics of the solutions. The near boundary expansion \eqref{asympSol} is the equivalent of the FG expansion \eqref{Fefferman-Graham} in the first order formulation of gravity.  The asymptotic expansion \eqref{asympSol} should still satisfy the constrain equation \eqref{cEquation}. As it was shown in \cite{Banados,Miskovic}, the constraint implies the hydrodynamic equations of motion for the energy momentum tensor and spin current, leaving the fields $e^a$ and $\omega^{ab}$ unconstrained, which will then be identified with the sources for $\tau_a$ and $\sigma_{ab}$. 
 
\subsection{Holographic Counterterm Action}

An important aspect of the holographic theory is renormalization. In order to identify the on-shell gravity action with the effective action of the field theory, we should first construct a finite action which at the same time preserve the boundary symmetries 
\begin{equation}\label{renAction1}
I_{\text{ren}}=\lim_{\epsilon \rightarrow 0} \left[ S_{\text{on-shell}}(\epsilon) - V(\epsilon) \right] \, ,   
\end{equation}
where $V(\epsilon)$ is the counterterm action and $\epsilon$ a cutoff for the FG coordinate. The renormalized action \eqref{renAction1} should have a well defined variational problem, so that, upon the holographic identification \eqref{fieldOnShell}, it becomes the generating function in the dual field theory. Following a procedure analogous to the one in \cite{Banados}, we obtain the following counterterm action 
\begin{align}\label{counterTerm}
\begin{split}
V &= 4 \kappa \epsilon_{abcd} \int_{M_4} \left[ (R^{ab} + e^a k^b) k^c e^d  \vphantom{\frac{\tilde e^a \tilde e^b \left(R^{cd}+ \frac{4}{3}\tilde e^c \tilde k^d   \right)}{2\epsilon}}  +\frac{\tilde e^a \tilde e^b \tilde e^c \tilde e^d}{6 \epsilon^2}  -\frac{\tilde e^a \tilde e^b \left(R^{cd}+ \frac{4}{3}\tilde e^c \tilde k^d   \right)}{2\epsilon}    \right]\, , 
\end{split}
\end{align}
with $R^{cd}= d \omega^{cd} + \omega^{c}_{\hphantom{c}b} \omega^{bd}$ the boundary field strength of the source $\omega^{ab}$. The first term of \eqref{counterTerm} is a Gibbons-Hawking term, while the last two terms cancel the divergences in $S_{\text{on-shell}}$. By varying $\delta I_{\text{ren}}$ and identifying it with the variation of  \eqref{fieldOnShell} the energy momentum and spin current three forms are found as \footnote{We note that the form of these currents correspond to a particular choice of finite counterterms given in (\ref{counterTerm}).}  
 \begin{equation}\label{t3}
 \tau_a = -8 \kappa \epsilon_{abcd} \left(R^{bc} + 2 e^b k^c \right)k^d \, ,
 \end{equation}
 \begin{equation}\label{s3}
 \sigma_{a b} =  16 \kappa \epsilon_{abcd} T^c k^d \, ,
 \end{equation}
with $T^c$ the torsion coming from the boundary sources $e^a$ and $\omega^{ab}$, see (\ref{TCD}). Following \cite{Banados,Miskovic} we obtain the corresponding conservation equations for $\tau_a$ and $\sigma_{ab}$ as, see section \ref{sec::Noether},  
\begin{align}\label{cons1}
\begin{split}
D \tau_a = I_a T^b \tau_b + \frac{1}{2} I_a R^{bc} \sigma_{bc}\, , \quad \quad \quad 
D\sigma_{ab} = e_a \tau_b - e_b \tau_a \, , 
\end{split}
\end{align}
These two equations are equivalent to the more familiar ones
\begin{align}\label{cons2}
\begin{split}
\nabla_\mu(|e| T^{\mu}_{\nu}) = \frac{|e|}{2} S^{\lambda}_{\hphantom{\lambda} \rho \sigma} R^{\rho \sigma}_{\hphantom{\mu \nu} \lambda \nu } \, , \quad \quad \quad
\nabla_\lambda \left(|e| S^\lambda_{\hphantom{\lambda} \rho \sigma}  \right) = \frac{|e|}{2} T_{[\rho \sigma]} \, ,
\end{split}
\end{align}
are the expected Ward identities for boundary diffeomorphism and local Lorentz invariance. The chosen gauge leaves not only these symmetries as residual boundary symmetries but also Weyl symmetry and a particular non-abelian symmetry are present \cite{Miskovic}. These last two symmetries become anomalous and it has been suggested in \cite{Banados} that the non-abelian anomaly could be related to a chiral anomaly through the antisymmetric part of the spin current. In appendix \ref{sec::Noether} we give a quick survey of all the symmetries and anomalies of the theory that survive the gauge fixing.

\subsection{Thermodynamic properties of the blackhole solutions}

In equilibrium we can Wick rotate the theory to Euclidian signature allowing us to identify the (grand-)canonical free energy $F_{\text{free}}$ as
\begin{align}\label{free1}
\beta F_{\text{free}} \equiv  \beta \int_{M_3 } \mathcal{F}_{\text{free}} =  I[e,\omega]_{\text{on-shell}} \, ,
\end{align}
with $\beta$ the inverse temperature which equals the length of the thermal cycle and $\mathcal{F}_{\text{free}}$ the free energy density integrated over the spatial boundary $M_3$. We are also interested in the thermal entropy $S_{\text{thermal}}$. For blackhole solutions the entropy can be computed as the Noether charge associated with diffeomorphisms\footnote{To be precise, here we consider the gauge invariant diffeomorphisms discussed below equation (\ref{dynamicalEq}).} and is found to be 
\begin{align}\label{entropyCS}
S_{\text{thermal}} \equiv \int_{M^h_3} \mathcal{S}_{\text{thermal}} = 4 \pi \int_{M^h_3} \epsilon_{A B C D F}  \left( \hat R^{C D} \hat e^F + \frac{1}{3} \hat e^C \hat e^D \hat e^F \right) n ^{AB} \, ,
\end{align}
where $M^h_3$ denotes the horizon manifold,  $n^{AB}\equiv D^{[A}\xi^{B]}$ is the binormal at the horizon, and $\xi^B$ the Killing vector generating the horizon.  We denote the entropy density by $\mathcal{S}_{\text{thermal}}$. Expression \eqref{entropyCS} is obtained from the general entropy formula \eqref{WaldEntropy} which is valid for a generic Lovelock gravity. This formula is the analogue of Wald's entropy formula \cite{WaldOriginal} for the first order formulation of gravity and its derivation is shown in appendix \ref{sec::Appendix::Entropy}. Our derivation closely follows \cite{Jacobson,Speziale} where a similar formula was derived for torsionless theories\footnote{There exist some subtitles for deriving a first law in the context of the first order formalism.}. The free energy and the entropy satisfy the Smarr relation and the first law of thermodynamics 
\begin{align}
\label{SmarrRel}
\mathcal{F}_{\text{free}} &= M_0 - T \mathcal{S}_{\text{thermal}} - \mu_I Q^I \, ,  \\
\label{FirstLawOne}
d\mathcal{F}_{\text{free}}&= -  \mathcal{S}_{\text{thermal}} dT - Q^I d\mu_I\, ,
\end{align}
with $M \equiv M_0 - \mu_I Q^I$ the mass of the blackhole. This mass  can be independently computed from the energy momentum tensor and the spin current as   
\begin{align}
\begin{split}\label{mass}
M= \int_{M_3} d^3 x \left[  n_\mu \xi_\nu T^{\mu \nu} + \frac{1}{2} n_\lambda S^\lambda_{\hphantom{\lambda} \mu \nu}\omega^{\mu \nu}_{\hphantom{\mu \nu} \alpha} \xi^\alpha  \right] \, ,
\end{split}
\end{align}
with $n_\mu$ the normal to the timelike direction, $\mu_I$ the spin sources, and $Q_I$ their associated charges. The derivation of the thermodynamic quantities presented this section --- as far as we know --- is novel. They form the basis of the derivation of the thermodynamic properties of the holographic spin fluids we discuss in the next three sections. 

\section{Generic holographic background}\label{sec::Hydro}

In this section we use the hydrodynamic and irreducible decompositions described in sections \ref{sec::SpinHydro} and \ref{sec::IrreSpin} to introduce an ansatz for $\{\hat e^M, \hat \omega^{MN} \}$ suitable for a blackhole solution. We will introduce this ansatz  in a form appropriate to perform the hydrodynamic expansion in (boundary) derivatives us to interpret the currents of the solution as those of a thermal conformal fluid. Just as in the derivation of the holographic currents above, we consider the following gauge 
\begin{align}\label{gaugeFixingFinal}
\begin{split}
\hat e^5 &= \frac{dr}{r g} +  \partial_\mu \left( \int \frac{dr}{rg} \right)dx^\mu  \, ,\\
\hat e^a_r & = \hat \omega^{a 5}_{\hphantom{a 5}r} = \hat \omega^{ab}_{\hphantom{ab}r} = \hat T^{A}_{\hphantom{A} r \mu} = F^{AB}_{\hphantom{AB} r \mu} = 0\, .
\end{split}
\end{align}
where $F^{AB}\equiv R^{AB} + e^B e^B$ is the Chern-Simons field strength. From section \ref{sec::Holography} we know that solving for $F^{AB}_{\hphantom{AB} r \mu}=0$ and $\hat T^A_{\hphantom{A} r \mu}=0$  is sufficient to determine to the radial dependence of the functions $\{ \hat e^a_\mu, \hat \omega^{a5}_{\hphantom{a5}\mu} , \hat \omega^{ab}_{\hphantom{ab}\mu} \} $ leaving a set of constraint equations for the integration constants and sources. The dynamical equations we need to solve are 
\begin{align}
\begin{split}\label{newEQM}
\partial_r \hat \omega^{ab}_{\hphantom{ab} \mu} &= 0 \, ,\\
\hat \omega^{a 5}_{\hphantom{a5}\mu} &= r g \partial_r \hat e^a_\mu \, , \\
\partial_r \left( r g \partial_r \hat e^a_\mu  \right) &= \frac{1}{rg} \hat e^a_\mu \, .
\end{split}
\end{align}
The system \eqref{newEQM} has a simple solution which we presented in section \ref{sec::Holography}. We will rewrite this solution in an alternative form by using the hydrodynamic ansatz presented in the rest of this section. This manipulation will allow us to solve with the remaining constraint equations more easily in terms of the hydrodynamic derivative expansion. This, in particular, entails a decomposition of the background ansatz in terms of  the fluid velocity $u^\mu$ and the projection operator $\Delta^\mu_\nu$.   

\subsection{Ansatz for the vielbein }

The one form $e^a$ can be  decomposed as, 
\begin{align}\label{vielbeinAnsatz}
\hat e^a &= \theta^a_\mu \left[ F(x,r) u^\mu u_\nu +  F_\sigma(x,r)  u^\mu \Delta^\sigma_\nu + \tilde F_\rho(x,r) \Delta^{\rho \mu} u_\nu + F_{\rho \sigma}(x,r) \Delta^{\rho \mu} \Delta^\sigma_\nu    \right] dx^\nu \, ,
\end{align}
where $\{F,F_{\sigma},\tilde F_\rho, F_{\rho \sigma} \}$ are functions of both the boundary and the holographic coordinate and $\theta^a$ is a boundary vielbein that is related to the vielbein $e^a$ in the previous section as  
\begin{align}\label{thetae}
e^a=\theta^a_\mu \left(u^\mu u_\nu + \Delta^\mu_\nu \right) dx^\nu \, .
\end{align}
The corresponding bulk metric reads 
\begin{align}
\begin{split}\label{metricAnsatz}
ds^2 &= \frac{dr^2}{r^2 g^2} + \left[ -\left(F^2 - \tilde F_\alpha \tilde F_\beta \Delta^{\alpha \beta} \right) u_\mu u_\nu + 2\left(\tilde F_\alpha F_{\beta \sigma}\Delta^{\alpha \beta} -F F_{\sigma}  \right) u_{\mu}\Delta^\sigma_{\nu} \right. \\ &\hphantom{=} \left. +  \left(F_{\alpha \rho}F_{\beta \sigma} \Delta^{\alpha \beta} - F_\rho F_\sigma   \right) \Delta^\rho_{\mu}\Delta^\sigma_{\nu} \right] dx^\mu dx^\nu \, ,
\end{split}
\end{align}
The AdS boundary conditions for the functions $\{g, F,F_\sigma,\tilde F_\sigma, F_{\rho \sigma} \}$  are such that 
\begin{align}\label{boundaryCondM}
\lim_{r \rightarrow \infty} ds^2 \sim  \frac{dr^2}{r^2} + r^2 \gamma_{\mu \nu} dx^\mu dx^\nu \, ,
\end{align}
where $\gamma_{\mu \nu}$ is the boundary metric coupled to the dual CFT. For simplicity we take $\gamma_{\mu \nu}= \eta_{\mu \nu}$ in the rest of the paper. It is now possible to write down an ansatz for the functions $\{g, F,F_\sigma,\tilde F_\sigma, F_{\rho \sigma} \}$ as an expansion in derivatives 
\begin{align}
 F(x,r)&= -r f(r) + \sum_{m=1}^{\infty} \sum_{i_m=1}^{I^s_m}\epsilon^{m} r f^{[m]}_{i_m}(r,x) s^{[m]}_{i_m}(x) \, ,\label{F1A}  \\  
F_\sigma(x,r)\Delta^{\sigma \mu} &= \sum_{m=1}^{\infty} \sum_{i_m=1}^{I^v_m}\epsilon^{m} r l^{[m]}_{i_m}(r,x) \upsilon^{[m] \mu}_{i_m}(x) \, , \label{F2A} \\
\tilde F_{\rho}(x,r) \Delta^{\rho \mu} &= \sum_{m=1}^{\infty} \sum_{i_m=1}^{I^v_m} \epsilon^{m} r \tilde  l^{[m]}_{i_m}(r,x) \upsilon^{[m]  \mu}_{i_m}(x) \, , \label{F3A} \\
F_{\rho \sigma}(x,r) \Delta^{\rho \mu} \Delta^{\sigma \nu} &= r h(r,x) \Delta^{\mu \nu} + \sum_{m=1}^{\infty} \sum_{i_m=1}^{I^t_m} \epsilon^{m} r h^{[m]}_{i_m}(r,x) t^{[m] \mu \nu}_{i_m}(x) \, , \label{F4A}
\end{align}
where $f$ a function of the radial coordinate\footnote{Using radial diffeomorphisms $f$ can be chosen to depend only on the holographic coordinate.}, $\{h,f^{[m]}_{i_m},l^{[m]}_{i_m},\tilde l^{[m]}_{i_m},h^{[m]}_{i_m}\}$ functions of both the holographic and the boundary coordinates\footnote{Dependence on the boundary coordinates in these functions arises from their temperature dependence.}, index $[m]$ indicates the order of appearance in the derivative expansion, $\epsilon$ is a book keeping parameter explicitly counting the number of derivatives. Here $s^{[m]}_{i_m}$,  
$\upsilon^{[m] \mu}_{i_m}$ and $t^{[m]\mu \nu}_{i_m}$ are, respectively, a scalar, a vector and a tensor from the corresponding $I^s_m$, $I^v_m$ and $I^t_m$ independent  quantities constructed from all available sources with $m$ number of derivatives. Although the ansatz in \eqref{F1A}-\eqref{F4A} is formal it is possible from \eqref{newEQM} to find a solution for all the functions in terms of the blackening factor $f$ as follows\footnote{This choice of ansatz with the corresponding AdS asymptotic behavior identifies the boundary source as in (\ref{thetae}).}:
\begin{align}
\begin{split}\label{solutionsGeneral}
g &= \frac{\sqrt{h_0+ (rf)^2}}{r (rf)'} \, , \\ 
h &= \frac{(h_0+h_1) (rf) + (h_0-h_1) \sqrt{h_0+ (rf)^2}}{2 r h_0} \, , \\
f^{[m]}_{i_m} &=  a^{[m]}_{i_m} \left( \frac{rf -\sqrt{h_0 + (rf)^2} }{2rh_0} \right) \, , \\
l^{[m]}_{i_m} &=  b^{[m]}_{i_m} \left( \frac{rf -\sqrt{h_0 + (rf)^2} }{2rh_0} \right) \, ,\\
\tilde l^{[m]}_{i_m} &=  c^{[m]}_{i_m} \left( \frac{rf -\sqrt{h_0 + (rf)^2} }{2rh_0} \right) \, , \\
h^{[m]}_{i_m} &=  d^{[m]}_{i_m} \left( \frac{rf -\sqrt{h_0 + (rf)^2} }{2rh_0} \right) \, , \\
\end{split}
\end{align}
with prime denoting a radial derivative, $\{h_0,a^{[m]}_{i_m},b^{[m]}_{i_m},c^{[m]}_{i_m},d^{[m]}_{i_m}\}$ integration constants that a priori are functions of the boundary coordinates, and where the boundary conditions \eqref{boundaryCondM} have already been implemented. As seen from \eqref{solutionsGeneral} we have rewritten the solution for the functions $\{g,h,f^{[m]}_{i_m},l^{[m]}_{i_m},\tilde l^{[m]}_{i_m},h^{[m]}_{i_m}\}$ as algebraic functions of $f$. This function $f$ is arbitrary and can be fixed via the remaining  diffeomorphisms of the holographic coordinate. 

It is instructive to look at the zeroth order in metric \eqref{metricAnsatz} 
\begin{align}\label{zeroBlackHole}
ds^2 = \frac{dr^2}{r^2 g^2} + r^2 \left(- f^2 u_\mu u_\nu + h^2 \Delta_{\mu \nu} \right) dx^\mu dx^\nu \, .
\end{align}
This corresponds to a non-extremal blackhole with a horizon at $r_h$ if the function $f$ satisfies $f^2(r_h)=0$ and $(f')^2(r_h)\neq 0$. Then the blackhole temperature $T(x)$ can be related to $h_0$ by
\begin{align}
h_0 = \left(2 \pi T \right)^2 \, . 
\end{align}

\subsection{Ansatz for the connection}

The connection $\omega^{a5}$ can be decomposed in a similar fashion 
\begin{align}
\hat \omega^{a5}=\theta^a_\mu \left[ K(x,r) u^\mu u_\nu +  K_\sigma(x,r)  u^\mu \Delta^\sigma_\nu + \tilde K_\rho(x,r) \Delta^{\rho \mu} u_\nu + K_{\rho \sigma}(x,r) \Delta^{\rho \mu} \Delta^\sigma_\nu    \right] dx^\nu 
\end{align}
where $\{K,K_\sigma,\tilde K_\rho,K_{\rho \sigma} \}$ are functions of the holographic and the boundary coordinates. From \eqref{newEQM} we see that these functions are solved in terms of  $\{F,F_{\sigma},\tilde F_\rho, F_{\rho \sigma} \}$  as
\begin{align}
\begin{split}
K &= (r g) F' \, ,\\
K_\sigma &= (r g) F'_\sigma \, , \\
\tilde K_\sigma &= (r g) \tilde F'_\sigma  \, , \\
K_{\rho \sigma} &= (r g) F'_{\rho \sigma} \, ,
\end{split}
\end{align}
where prime denotes derivative with respect to $r$. From \eqref{newEQM} it also follows that the connection $\hat \omega^{ab}$ should be independent of the holographic coordinate can only depend on the boundary coordinates. These, then, corresponds to the external spin sources $\omega^{ab}$ in dual field theory and decomposed according to \eqref{decompConnection}.

\section{Solutions dual to zeroth order hydrodynamics}
\label{sec::ZerothSolution}

We first consider the zeroth order in the derivative expansion where all sources are taken to be constant. Using  \eqref{solutionsGeneral} we reduce the equations of motion to a set of algebraic constraints. Before attempting to solve this set of constraints, we will first limit the space of solutions by analyzing the behavior of the Ricci scalar $\mathcal{R}$ near the horizon and demanding that it stays finite. At the horizon the Ricci scalar behaves as 
\begin{align}\label{RicciH0}
R(r_h+ \epsilon) =  \frac{1}{\epsilon} \left[ 4 \left(\mathcal{A}_1^\mu \mathcal{A}_2^\nu - \mathcal{V}^\mu_1 \mathcal{V}^\nu_2 \right)\gamma_{\mu \nu} - 3 \left( h_0- h_1 \right) \right] + \mathcal{O}\left( \epsilon^0 \right) \, .
\end{align}
A particular solution reads
\vspace{2mm}
\begin{align}
h_1 &= - h_0 \, , \label{RicciConstraints} \\
\mathcal{A}^\mu _1 \mathcal{A}^\nu_2 \gamma_{\mu \nu} &=\mathcal{V}^\mu_1 \mathcal{V}^\nu_{2} \gamma_{\mu \nu} \, . \label{RicciConstraints2} 
\end{align}
Clearly \eqref{RicciConstraints}-\eqref{RicciConstraints2} is not the most general regular solution, but there is another physical reason to require this: we showed that this solution is equivalent to the first law \eqref{FirstLawOne} of blackhole thermodynamics. In appendix \ref{sec::Appendix::SingularSolutions} we discuss some solutions which do not satisfy them. Here we will restrict our attention to blackholes with regular thermodynamics, hence require \eqref{RicciConstraints}-\eqref{RicciConstraints2}. 

We are then left with 9 independent algebraic constraint equations: two linear, three quadratic and four cubic constraint equations on the spin sources. The linear constraints
\begin{align}\label{linearConst1}
\mu_V &= 0 \, \\ 
\mathcal{V}^\mu_1 &= 0 \, \label{linearConst2} 
\end{align}
set two of the spin sources to zero. Using this, the three quadratic constraints are  
\begin{align}
\mathcal{A}_2^\alpha  u^\beta  \epsilon_{\alpha \beta \sigma [\mu} \mathcal{H}^\sigma_{\nu]} &= 0 \, , \label{quadConst1}\\
\left(\mathcal{C}_{\mu \alpha} + 4 \mu_A \gamma_{\mu \alpha} \right) \mathcal{A}^\alpha_2 +  \mathcal{H}_{\mu \alpha} \mathcal{V}^\alpha_2 + \epsilon_{\mu \alpha \beta \sigma} u^\alpha \left( \mathcal{A}_2^\beta \mathcal{V}^\sigma_2 + C^\beta_\rho C^{\rho \sigma}\right) &=0 \, , \\ 
\mathcal{A}^\alpha_1 \mathcal{A}^\beta_2 \gamma_{\alpha \beta} &= 0 \, ,
\end{align}
while the four cubic constraints become 
\begin{align}
\mathcal{H}_{\alpha \beta} \mathcal{A}^\alpha_1 \mathcal{A}^\beta_1 &= 0 \, , \\ 
\mathcal{H}_{\alpha \beta}\left(\mathcal{C}^\beta_\mu \mathcal{C}^{\mu \alpha} - \mathcal{H}^\beta_\mu \mathcal{H}^{\mu \alpha}+\mathcal{V}^\alpha_2 \mathcal{V}^\beta_2 - 3 \mathcal{A}^\alpha_2 \mathcal{A}^\beta_2 - \mathcal{C}^{\alpha \beta}\mu_A \right)  + 2 \mathcal{A}^\alpha_2 \mathcal{V}^\beta_2 \left( \mu_A \gamma_{\alpha \beta} + \mathcal{C}_{\alpha \beta} \right) &=0 \, ,  \\ \nonumber \mathcal{H}_{\alpha \beta} \left( \mathcal{H}^\beta_\mu \mathcal{V}^\beta_2 - \mathcal{C}^\beta_\mu \mathcal{A}^\alpha_2 + 2 \mu_A \delta^\beta_\mu \mathcal{A}^\alpha_2 \right) - 2\mathcal{A}_2^\alpha \mathcal{A}_2^\beta \mathcal{V}_2^\nu \gamma_{\alpha [\beta} \gamma_{\mu] \nu} + 2 u^\alpha \mathcal{A}^\beta_2 \mathcal{V}^\rho_2 \epsilon_{\alpha \beta \sigma [\mu} \mathcal{H}^\sigma_{\rho]} \\ 
 + u^\alpha \epsilon_{\alpha \rho \sigma \lambda} \mathcal{C}^{\nu \rho} \mathcal{H}^\lambda_\nu \mathcal{H}^\sigma_\mu + u^\alpha \epsilon_{\mu \alpha \rho \sigma} \mathcal{C}^\sigma_\nu \mathcal{A}^\nu_2 \mathcal{A}^\rho_2 &= 0 \, , \\
\nonumber \mathcal{A}^\alpha_1 \left( \mathcal{H}_{\alpha \beta} \mathcal{C}^\beta_\mu -  \gamma_{\alpha \mu} \left( \mathcal{H}_{\rho \sigma} \mathcal{C}^{\rho \sigma} -  \mathcal{A}^\rho_2 \mathcal{V}_{2 \rho} \right) + 2 \mathcal{C}^\beta_\alpha \mathcal{H}_{\mu \beta} - \mathcal{A}_{2 \mu} \mathcal{V}_{2 \alpha} + 4 \mathcal{A}_{2 \alpha} \mathcal{V}_{2 \mu} - 3 \mu_A \mathcal{H}_{\mu \alpha}  \right) \\
\nonumber + \epsilon_{\mu \alpha \beta \sigma} u^\alpha \left( \mathcal{C}^{\beta}_{\rho}\mathcal{H}_\rho^\nu \mathcal{H}^\sigma_\nu - 4 \mathcal{A}^\beta_1 \mathcal{C}^\sigma_\rho \mathcal{A}^\rho_2 - \mathcal{A}^\beta_1 \mathcal{H}^\sigma_\rho \mathcal{V}^\rho_2 - 5 \mu_A \mathcal{A}^\beta_1 \mathcal{A}^\sigma_2 \right) \\ 
+ \epsilon_{\alpha \beta \rho \sigma} u^\alpha  \left( \mathcal{C}^\beta_\lambda \mathcal{H}^{\lambda \sigma} \mathcal{H}^\rho_\mu + \mathcal{A}^\beta_1 \mathcal{A}^\rho_2 \mathcal{C}^\sigma_\mu - \mathcal{A}^\beta_1 \mathcal{V}^\rho_2 \mathcal{H}^\sigma_\mu  \right)   &=0  \label{cubicConst4}\, .
\end{align}
Independence among the subset of non-vanishing sources requires the rest to vanish. We find two classes of solutions satisfying this requirement, 
\begin{itemize}
\item Vanishing $\{ \mathcal{H}^{\mu \nu}, \mu_V, \mathcal{V}_1^\mu, \mathcal{A}^\mu_2 \}$ and independent sources $\{\mathcal{C}^{\mu \nu}, \mathcal{V}^\mu_2,\mu_A , \mathcal{A}^\mu_1 \}$. We will refer to this solution as the Scalar-Vector-Tensor solution. 
\item Vanishing  $\{ \mathcal{H}^{\mu \nu}, \mathcal{C}^{\mu \nu}, \mu_V, \mathcal{V}_1^\mu, \mathcal{V}_2^\mu, \mu_A,\mathcal{A}^\mu_1 \}$ and the single independent source $\mathcal{A}^\mu_2$. We will refer to this solution as the Axial solution.  
\end{itemize}
Both classes satisfy the first law \eqref{FirstLawOne}. Below we analyze these two solutions by computing their zeroth order constitutive relations and determining their thermodynamic behavior. 

\subsection{Scalar-Vector-Tensor solution}

For sake of clarity we present the explicit form of the metric and the torsion with non-vanishing scalar-vector-tensor in a convenient gauge\footnote{This gauge puts the blackhole solution in the familiar form in higher derivative gravity, e.g. the Gauss-Bonnet gravity with the Gauss-Bonnet coupling set to $\lambda_{GB} = 1/4$, see for example \cite{Ross, Liu}.}  $f^2 = 1 - \frac{r^2_h}{r^2}$.
\begin{align}
\begin{split}\label{solutionGeneral0}
ds^2 &= - \frac{dr^2}{r^2 g^2} + r^2 \left(-f^2 u_\mu u_\nu + h^2 \Delta_{\mu \nu}  \right) dx^\mu dx^\nu \, , \\
& f^2  = 1 - \frac{r^2_h}{r^2} \, , \\
& g^2 = \left( 1 - \frac{r^2_h}{r^2} \right) \left(1 - \frac{r^2_h-4 \pi^2 T^2 }{r^2} \right)  \, , \\
& h^2 = 1 - \frac{r^2_h - 4 \pi^2 T^2}{r^2} \, , \\
T^a &=  r h \theta^a_\mu \left[  \epsilon^{\mu}_{\hphantom{\mu} \nu \rho \alpha} u^\nu \left( \mathcal{C}^\alpha_\sigma + \mathcal{A}^\alpha_1 u_\sigma - \mu_A \delta^\alpha_\sigma  \right) - \mathcal{V}_{2 \rho} \Delta^{\mu}_{\sigma}    \right]dx^\rho \wedge dx^\sigma   \, ,\\
T^5 &= 0 \, .
\end{split}
\end{align}
The energy momentum tensor of the holographic dual fluid follows from \eqref{t3} and \eqref{ts3} as 
\begin{align}\label{energyMomentum0T}
\begin{split}
 T^{\mu \nu} &=  32 \pi^4 \kappa T^4 \left[\left(1 + \frac{C^\rho_\sigma C^\sigma_\rho-2\mu^2}{4 \pi^2 T^2} \right) \left(4 u^\mu u^\nu + \gamma^{\mu \nu} \right)  + \frac{u^\mu \left(2 \mu_A \delta^\nu_\rho - C^\nu_\rho  \right) \mathcal{A}^\rho_1}{2 \pi^2 T^2}   \right. \\  &\hphantom{=} \hphantom{32 \pi^2 \kappa T^4 [[[[}\left. - \frac{u^\mu u^\nu \left(\mathcal{V}^2_2 + C^{\rho \sigma}C_{\rho \sigma} \right)  + \mathcal{V}^\mu_2 \mathcal{V}^\nu_2 - \mu_A C^{\mu \nu} + C^\mu_\rho C^{\rho \nu}}{2 \pi^2 T^2} \right. \\  &\hphantom{=} \hphantom{32 \pi^2 \kappa T^4 [[[[} \left. + \frac{ \epsilon^{\mu \nu \rho \sigma} \left(\mu_A u_\rho \gamma_{\sigma \lambda} +  u_\rho C_{\sigma \lambda} \right) \mathcal{V}_{2}^{\lambda} - u^\mu \epsilon^\nu_{\hphantom{\nu} \alpha \beta \lambda} \mathcal{A}^\alpha_1 \mathcal{V}^\beta_2 u^\lambda  }{2 \pi^2 T^2}  \right]\, ,
\end{split}
\end{align}
We read off the physical quantities by projections of this onto different components using \eqref{projectionsT}: 
\begin{align}
\begin{split}\label{energyMProjections0T}
\varepsilon &= 96 \pi^4 \kappa T^4 \left(1 + \frac{\mathcal{C}^{\alpha \beta} \mathcal{C}_{\alpha \beta} - 2 \mathcal{V}_2^2 - 6 \mu^2_A}{12 \pi^2 T^2} \right) \, , \\
p &= -32 \pi^4 \kappa T^4 \left(  1 + \frac{\mathcal{C}^{\alpha \beta} \mathcal{C}_{\alpha \beta} - 2 \mathcal{V}_2^2 - 6 \mu^2_A}{12 \pi^2 T^2}\right) \, , \\
\bar q^\mu &= 0 \, , \\
q^\mu &= -16 \pi^2 \kappa T^2 \left( \mathcal{C}^\mu_\alpha \mathcal{A}^\alpha_1 - 2 \mu_A \mathcal{A}^\mu_1 + \epsilon^{\mu \alpha \rho \sigma} u_\alpha \mathcal{A}_{1 \rho} \mathcal{V}_{2 \sigma}\right) \, , \\
\pi^{\mu \nu} &= -16 \pi^2 \kappa T^2 \left(  C^{\mu \alpha} C^\nu_\alpha + \mathcal{V}^\mu_2 \mathcal{V}^\nu_2 - \frac{1}{3} \left( \mathcal{C}^\alpha_\beta \mathcal{C}^\beta_\alpha + \mathcal{V}^2_2  \right) \Delta^{\mu \nu}  - \mu_A \mathcal{C}^{\mu \nu}  \right) \, , \\
\tau^{\mu \nu} &=  16 \pi^2 \kappa T^2 u_\rho \mathcal{V}^\beta_2 \epsilon^{\mu \nu \rho \sigma } \left( \mathcal{C}^\sigma_\beta + \mu_A \delta^\sigma_\beta  \right)\, . 
\end{split}
\end{align}
The energy momentum is traceless and satisfies the usual equation of state $p=-\frac{\varepsilon}{3}$ for a conformal fluid. There exists a non-vanishing shear (traceless symmetric) component when  $C^{\mu \nu}$ or $\mathcal{V}_2$ are non vanishing, Also we observe from $\Delta q^\mu \neq 0$ and $\tau^{\mu \nu} \neq 0$ that the energy momentum tensor is not symmetric unless $\{\mathcal{A}_1, \mathcal{V}_2 \}$ or $\{\mathcal{A}_1, \mathcal{C}^{\mu \nu}, \mu_A \}$ vanish. Finally, we observe several novel transport coefficients associated to the spin sources appear in \eqref{energyMomentum0T}. However we will not discuss them in detail in this work, and leave a detailed study of spin-related transport coefficients to future work.

The spin current in this holographic fluid follows from \eqref{s3} and \eqref{ts3} as 
\begin{align}\label{S1}
 S^{\lambda \mu \nu} &= 32 \pi^2 \kappa T^2 \left[ \mu_A \epsilon^{\lambda \mu \nu \alpha} u_\alpha - u_\alpha  \epsilon^{\lambda \alpha \beta [\mu} \left( \mathcal{C}^{\nu]}_\beta + u^{\nu]} \mathcal{A}_{1\beta} \right) + \Delta^{\lambda [\mu} \mathcal{V}^{\nu]}_2 + 2 u^\lambda u^{[\mu} \mathcal{V}^{\nu]}_2  \right] \, ,
\end{align}
which can be decomposed into irreducible scalar, vector and tensor parts as (see section \ref{sec::hydro})
\begin{align}
\begin{split}\label{spinProj0}
\rho_V &= 0 \, ,\\
\rho_A &= 32 \pi^2 \kappa T^2 \mu_A \, , \\
n^\mu_V &= -32 \pi^2 \kappa T^2 \mathcal{V}^\mu_2 \, , \\
\bar n^\mu_V &= 16 \pi^2 \kappa T^2 \mathcal{V}^\mu_2 \, ,\\
n^\mu_A &= 0 \, ,\\
\bar n^\mu_A &= -16 \pi^2 \kappa T^2 \mathcal{A}^\mu_1 \, ,\\
N^\lambda_\kappa &= 0 \, ,\\
\bar N^\lambda_\kappa &= -16 \pi^2 \kappa T^2 C^\lambda_\kappa \, .
\end{split}
\end{align}
We also present the corresponding vector components in the ``irreducible" decomposition in section \ref{sec::hydro}
\begin{align}
\begin{split}\label{spinProj02}
J^\mu_V &= 0\, , \\
\bar J^\mu_V &= 32 \pi^2 \kappa T^2  \mathcal{V}^\mu_2 \, ,\\
J^\mu_A &= - 32 \pi^2 \kappa T^2 \mu_A u^\mu - \frac{32 \pi^2 \kappa T^2}{3} \mathcal{A}^\mu_1 \, , \\
\bar J_A &= - \frac{32 \pi^2 \kappa T^2}{3} \mathcal{A}^\mu_1 \, .
\end{split}
\end{align}
All non vanishing currents in \eqref{spinProj0} and \eqref{spinProj02} are linear in the sources. There is a single vector current of irreducible vector-tensor type while there is a single axial current of the hydrodynamic type. 

An interesting observation here --- which confirms and forms the basis for our discussion in the Introduction --- is that, there exists a non-trivial spin current, due to the presence of non-trivial spin sources,  even when the energy-momentum tensor is symmetric. This is because $S^{\lambda \mu \nu}$ in (\ref{S1}) is non-trivial both when $\mathcal{A}_1 = \mathcal{V}_2 =0 $ or $\mathcal{A}_1= \mathcal{C}^{\mu \nu}=  \mu_A = 0$. 

Finally the thermodynamic potentials which correspond to the mass $M_0$, the axial charge $Q_A$, the vector charge $Q^\mu_{\mathcal{V}_2}$, the symmetric and traceless tensor charge $Q^{\mu \nu}_C$,  the free energy density, and the entropy density of the holographic fluid are given by 
\begin{align}
\begin{split}\label{thermPotGen}
M_0 &= 96 \pi^4 \kappa T^4 \left( 1 - \frac{3 \mu^2_{\text{eff}} }{2 \pi^2 T^2} \right)   \, , \\
Q_A &= -96 \pi^2 \kappa T^2 \mu_A \, , \\
Q^\mu_{\mathcal{V}_2} &=  -32 \pi^2 \kappa T^2 \mathcal{V}^\mu_2   \, , \\
Q^{\mu \nu}_C &=  16 \pi^2 \kappa T^2 C^{\mu \nu}  \, , \\
\mathcal{F}_{\text{free}} &= - 32 \pi^4 \kappa T^4 \left( 1 - \frac{3 \mu^2_{\text{eff}}}{2 \pi^2 T^2} \right)   \, ,\\ 
\mathcal{S}_{\text{thermal}} &= 128 \pi^4 \kappa T^3 \left(1 - \frac{3 \mu^2_{\text{eff}}}{4 \pi^2 T^2}\right) \, ,
\end{split}
\end{align}
where we defined an effective chemical potential $\mu_{\text{eff}}$ 
\begin{align}\label{mueff}
\mu^2_{\text{eff}} \equiv \mu_A^2 + \frac{1}{3} \mathcal{V}^2_2  - \frac{1}{6} C^\rho_\sigma C^\sigma_\rho = \frac{2 Q^2_A - 12 Q^2_C + 3 Q^2_{\mathcal{V}_2}}{18432 \pi^4 \kappa T^4} \, ,
\end{align}
which enters in the free energy and the entropy density. We note that the energy of the fluid, (\ref{energyMProjections0T}), and the total mass of the blackhole agrees $\varepsilon = M = M_0 - \mu_I Q^I$, see (\ref{mass}). 

 Positivity of the thermal entropy requires $\mu_{\text{eff}}<\frac{4\pi^2 T^2}{3}$ for $\kappa>0$ and $\mu_{\text{eff}}>\frac{4\pi^2 T^2}{3}$ for $\kappa<0$. As the potentials \eqref{thermPotGen} are obtained from \eqref{free1},\eqref{entropyCS}, and \eqref{mass} they should satisfy the first law of thermodynamcis. This is verified straightforwardly
\begin{align}
\begin{split}
d \mathcal{F}_{\text{free}} &= - \mathcal{S}_{\text{thermal}} dT - Q_A d \mu_A - Q_{\mathcal{V}_2}^\mu d \mathcal{V}_{2 \mu} - Q^{\mu \nu}_C dC_{\mu \nu} \, .
\end{split}
\end{align}
The thermodynamic stability of the solution requires  positivity of the specific heat $c$, and the chemical susceptibilities $\chi$, 
\begin{align}\label{stability}
\begin{split}
C_V \equiv - T \frac{\partial^2 \mathcal{F}_{\text{free}}}{\partial T^2} =  384 \pi^4 \kappa T^3 \left(1 -\frac{\mu^2_{\text{eff}}}{ 4 \pi^2 T^2} \right) & > 0 \, ,\\ 
\chi_A  \equiv \frac{\partial Q_A}{\partial \mu_A} = -96 \pi^2 \kappa T^2 &>0 \, , \\
\chi_{\mathcal{V}_2}^{\mu \nu} \equiv \frac{\partial Q_{\mathcal{V}_2}^\mu}{\partial \mathcal{V}_{2 \nu }} =  - 32 \pi^2 \kappa T^2 \Delta^{\mu \nu} &> 0 \, , \\
\chi^{\alpha \beta \rho \sigma}_{C} \equiv \frac{\partial Q^{\alpha \beta}_C}{\partial C_{\rho \sigma}} = 16 \pi^2 \kappa T^2 \Delta^{\alpha \rho} \Delta^{\beta \sigma} &> 0 \, .
\end{split}
\end{align}
To analyze \eqref{stability} we need to distinguish the two cases: $\kappa>0$ and $\kappa<0$. For $\kappa>0$ we first observe that the positivity of the susceptibilities require setting $\mu_A = \mathcal{V}_2 = 0$. Then, from (\ref{mueff}) we find $\mu_{eff}^2 > 0$ hence there are no further conditions that arise from positivity of the specific heat or positivity of the entropy. We conclude that, for $\kappa>0$ the black hole with a tensor source is thermodynamically stable. For  $\kappa<0$  on the other hand positive susceptibilities require vanishing of the tensor source. From positivity of the specific heat then we obtain an upper bound on the temperature $ T < \mu_{eff}/2\pi$. Dynamical stability of the solutions is another question which can be settled by considering perturbations around the background and calculating the quasi-normal modes. We will not investigate dynamical stability of the solutions in this paper. 
\subsection{Single axial solution}

Now we consider the second class of solutions to the constraint equations, outlined in the beginning of this section: the blackholes with a single axial charge. The explicit forms of the metric and the torsion, choosing once again the gauge $f^2 = 1 -\frac{r^2_h}{r^2}$,  read
\begin{align}
\begin{split}
ds^2 &= - \frac{dr^2}{r^2 g^2} + r^2 \left(-f^2 u_\mu u_\nu + h^2 \Delta_{\mu \nu}  \right) dx^\mu dx^\nu \, , \\
& f^2  = 1 - \frac{r^2_h}{r^2} \, , \\
& g^2 = \left( 1 - \frac{r^2_h}{r^2} \right) \left(1 - \frac{r^2_h-4 \pi^2 T^2 }{r^2} \right)  \, \\
& h^2 = 1 - \frac{r^2_h - 4 \pi^2 T^2}{r^2} \, , \\
T^a &= r \theta^a_\mu u^\alpha \mathcal{A}^\beta_2  \left(f \gamma^{\lambda \mu} \epsilon_{\alpha \beta \lambda  \sigma} u_\rho   -h u^\mu \epsilon_{\alpha \beta \rho \sigma}  \right) dx^\rho \wedge dx^\sigma  \, ,\\
T^5 &= 0 \, ,
\end{split}
\end{align}
We find the energy-momentum tensor from \eqref{t3} and \eqref{ts3} as 
\begin{align}
\begin{split}
T^{\mu \nu} &= 32 \pi^4 \kappa T^4 \left[ 4 u^\mu u^\nu + \gamma^{\mu \nu} + \frac{\mathcal{A}^2_2 u^\mu u^\nu + \mathcal{A}_2^\mu \mathcal{A}^\nu_2 }{2 \pi^2 T^2} \right]\, ,
\end{split}
\end{align}
with the following non-vanishing hydrodynamic projections, energy, pressure and shear (symmetric traceless) component, see (\ref{compTmn}): 
\begin{align}
\begin{split}\label{projectTA2}
\varepsilon &= 96 \pi^4 \kappa T^4 \left(1 + \frac{\mathcal{A}^2_2}{6 \pi^2 T^2} \right) \, , \\
p &=- 32 \pi^4 \kappa T^4 \left(1 + \frac{\mathcal{A}^2_2}{6 \pi^2 T^2}  \right)\, , \\
\pi^{\mu \nu} &=  16 \pi^2 \kappa T^2 \left( \Delta^\mu_{(\rho}\Delta^\nu_{\sigma)} - \frac{1}{3} \Delta^{\mu \nu} \Delta_{\rho \sigma}  \right)\mathcal{A}^\rho_2 \mathcal{A}^\sigma_2 \, . \\
\end{split}
\end{align}
We observe that the energy momentum tensor for the single axial fluid is traceless and symmetric with a shear component proportional to the tensor $\mathcal{A}^\rho_2 \mathcal{A}^\sigma_2$. The spin current follows from \eqref{s3} and \eqref{ts3} as 
\begin{align}
\begin{split}\label{currentsA2}
S^{\lambda \mu \nu} &=- 32 \pi^2 \kappa T^2  \left( u^\lambda \epsilon^{\mu \nu \alpha \beta}+ \epsilon^{\lambda \alpha \beta[\mu} u^{\nu]}  \right) u_\alpha \mathcal{A}_{2 \beta}\, ,
\end{split}
\end{align}
whose only non-vanishing hydrodynamic projections are the two axial hydrodynamic currents given by 
\begin{align}
\begin{split}
n_A^\mu &= -32 \pi^2 \kappa T^2 \mathcal{A}^\mu_2  \, \\
\bar n_A^\mu &= -16 \pi^2 \kappa T^2 \mathcal{A}^\mu_2 \, .
\end{split}
\end{align}
As in the previous example we note that there exists a non-trivial spin current sourced by the torsion even though the energy-momentum tensor is symmetric. Contrary to the scalar-vector-tensor solution there is no naturally preferred decomposition of the axial current and either the hydrodynamic or the irreducible decomposition can be used.

As for the thermodynamics, there is an axial charge $Q^\mu_{\mathcal{A}_2}$ associated to $\mathcal{A}_2$ as shown in the thermodynamic potentials computed in \eqref{potA2}.
\begin{align}
\begin{split}\label{potA2}
M_0 &= 96 \pi^4 \kappa T^4 \left( 1 + \frac{\mathcal{A}^2_2}{2 \pi^2 T^2} \right)   \, , \\
Q^\mu_{\mathcal{A}_2} &=  32 \pi^2 \kappa T^2 \mathcal{A}^\mu_2   \, , \\
\mathcal{F}_{\text{free}} &= - 32 \pi^4 \kappa T^4 \left( 1 + \frac{\mathcal{A}^2_2}{2 \pi^2 T^2} \right)   \, ,\\ 
\mathcal{S}_{\text{thermal}} &= 128 \pi^4 \kappa T^3 \left(1 + \frac{\mathcal{A}^2_2}{4 \pi^2 T^2} \right) \, .
\end{split}
\end{align}
We again observe that the energy in (\ref{projectTA2}) agrees with the total mass in (\ref{mass}). We also note that positivity of thermal entropy automatically discards the case $\kappa<0$. The thermodynamic potentials in \eqref{potA2} satisfy the first law in the form 
\begin{align}
\begin{split}\label{FirstLawA2}
d \mathcal{F}_{\text{free}}= -  \mathcal{S}_{\text{thermal}} dT - Q^\mu_{\mathcal{A}_2} d \mathcal{A}^\mu_2 \, .
\end{split}
\end{align}
The stability of the solution is determined from the conditions on the specific heat and the susceptibilities 
\begin{align}
\begin{split}
C_V \equiv - T \frac{\partial^2 \mathcal{F}_{\text{free}}}{\partial T^2} =  384 \pi^4 \kappa T^3 \left(1 +\frac{\mathcal{A}^2_2}{ 12 \pi^2 T^2} \right) & > 0 \, ,\\ 
\chi^{\mu \nu}_{\mathcal{A}_2}  \equiv \frac{\partial Q^\mu_{\mathcal{A}_2}}{\partial \mathcal{A}_{2 \nu}} =32 \pi^2 \kappa T^2 \Delta^{\mu \nu} &>0 \, , \\
\end{split}
\end{align}
We observe that the thermodynamic stability is guaranteed for $\kappa>0$.
%
%
%

\section{Solutions dual to first order hydrodynamics}
\label{sec::FirstSolution}

We now promote the hydrodynamic variables $u^\mu$ and $T$ and the spin sources $\omega^{ab}_\mu$ to slowly varying functions of the boundary coordinates to study hydrodynamic expansion at first order in derivatives.  For this purpose we consider the generic solution in \eqref{solutionsGeneral} to solve the constraint equations \eqref{cEquation} up to first order in the derivative expansion. For simplicity we explicitly treat only two cases: non vanishing $\mu_A$ and  non vanishing $\mathcal{V}^\mu_2$. This is sufficiently rich to explore the spin dependent transport in out holographic model. For all cases we can choose the gauge $f^2= 1 - \frac{r^2_h}{r^2}$ as before. In this gauge the generic solution for the metric at first order can be written as    
\begin{align}
\begin{split}
ds^2 &=   \frac{dr^2}{r^2 g^2}+ r^2 \left[- f^2 u_\mu u_\nu + h^2 \Delta_{\mu \nu} +j \left( 2 h c_I \, u_\mu v^I_\nu - 2 f b_I \, u_\mu v^I_\nu + h \, d_I t_{\mu \nu}^I \right)       \right] dx^\mu dx^\nu  \, ,  \\ 
& f^2 = 1 - \frac{r^2_h}{r^2} \, ,\\
& g^2 = \left( 1 - \frac{r^2_h}{r^2} \right) \left(1 - \frac{r^2_h-4 \pi^2 T^2 }{r^2} \right)  \, , \\
& h^2 = 1 - \frac{r^2_h - 4 \pi^2 T^2}{r^2} \, , \\
& j \equiv   \frac{1}{2 h_0} \left[\sqrt{1  - \frac{r^2_h}{r^2}} - \sqrt{1 - \frac{r^2_h - 4 \pi^2 T^2}{r^2}}  \right] \, ,
\end{split}
\end{align}
where $\{c_I v^I_\mu, b_I v^I_\mu \}$ denoting two distinct combinations of linearly independent vectors with a single derivative of any of the hydrodynamic ($u^\mu$, $T$) or spin sources ($\omega^{ab}_\mu$), and $\{d_I t^I_{\mu \nu} \}$ the same for linearly independent tensors. To obtain the solution we used the regularity of the metric determinant to set $F=-r f$ in the metric ansatz \eqref{metricAnsatz} and \eqref{F1A}. The corresponding torsion at first order two form for non-vanishing $\{\mu_A, \mathcal{V}^\mu_2, \mathcal{C}^{\mu \nu}, \mathcal{A}^\mu_2\}$ is given by 
\begin{align}
\begin{split}
T^a &= \theta^a_\alpha \left[ r \left( -\epsilon^{\alpha \beta}_{\hphantom{\alpha \beta} \rho \sigma} u^\rho \left(\mu_A \delta^\sigma_\mu + \mathcal{C}^\sigma_\mu   \right) - 2 \mathcal{V}^{[\alpha}_2 \Delta^{\beta]}_{ \mu}   +  2u^{[\alpha} \epsilon^{\beta]}_{\hphantom{\beta} \rho \sigma \mu} u^\rho \mathcal{A}^\sigma_2   \right) \left(-f u_\beta u_\nu + h \Delta_{\beta \nu} \vphantom{ \mathcal{V}^{[\alpha}_2 } \right. \right. \\ &\hphantom{=} \hphantom{\theta^a_\alpha [ raaaaa}\left. \left. \vphantom{ \mathcal{V}^{[\alpha}_2 }  + j u_\beta  b_I v^I_\nu + j u_\nu  c_I v^I_\beta + j d_I t^I_{\beta \mu }   \right) + \frac{\Delta^\alpha_\nu \partial_\mu h_0}{2r h}  + 4 h_0 r j u_{(\mu} \partial_\nu u_{\alpha)}  \right. \\ &\hphantom{=} \hphantom{\theta^a[aaaa} \left.    -  \frac{  h u_\alpha u_\mu- f \Delta_{\alpha \mu} + j \left(u_\mu  c_I v^I_\alpha + u_\alpha b_I v^I_\mu  \right) + d_I t^I_{\mu \alpha}}{2 r h \left( f+ h \right) } \partial_\nu h_0 \right] dx^\mu \wedge dx^\nu  \, , \\
T^5 &= \frac{1}{2} \left[ d_I t^I_{\mu \nu}+  u_\mu \left(b_I - d_I)I \right)v^I_\nu  \right] dx^\mu \wedge dx^\nu \, .
\end{split}
\end{align}
It is convenient to decompose the gradient of the four velocity as 
\begin{align}
\begin{split}\label{uDecomp}
\partial_\mu u_\nu &= \frac{1}{3} \Theta \Delta_{\mu \nu} + \sigma_{\mu \nu} - u_\mu a_\nu - \frac{1}{2} \epsilon_{\mu \nu \rho \sigma} u^\rho \omega^\sigma \, ,
\end{split}
\end{align}
where the compressibility $\Theta$, acceleration $a_\mu$, vorticity $\omega_\mu$, and shear tensor $\sigma_{\mu \nu}$ are defined as the projections 
\begin{align}
\begin{split}\label{velocityDecomp}
\Theta &\equiv \partial_\lambda u^\lambda  \, ,\\
a^\mu &\equiv u^\lambda \partial_\lambda u^\mu \, ,\\
\omega^\mu &\equiv \epsilon^{\mu \nu \alpha \beta} u_\nu \partial_\alpha u_\beta \, , \\
\sigma^{\mu \nu} &\equiv \left(\Delta^\mu_{(\rho} \Delta^\nu_{\sigma)}- \frac{1}{3} \Delta^{\mu \nu} \Delta_{\rho \sigma} \right) \partial^\rho u^\sigma \, .
\end{split}
\end{align}
Using the decomposition \eqref{uDecomp} the condition for regularity of the Ricci scalar at the horizon can be written as 
\begin{align}
\begin{split}
d_I  t^{I \mu}_{\hphantom{\mu} \mu} = -4 a_\mu \mathcal{V}^\mu_2 - 2 \omega_\mu \mathcal{A}^\mu_2  \, .
\end{split}
\end{align}
Below we derive the constraint equations to be satisfied by the sources and he integration constants $\{b_I v^I_\mu,c_I v^I_\mu,d_I t^I_{\mu \nu} \}$ for the particular solutions. 

\subsection{Non vanishing $\mu_A$}

For the specific solution with only $\mu_A$ non-vanishing we obtain the following constraints: Three scalar equations 
\begin{align}
\begin{split}\label{case1eq1}
\left( 4 \pi^2 T^2 - \mu_A^2 \right) \Theta &= 0 \, ,\\
12 \mu_A u^\alpha \partial_\alpha \mu_A + 2 \Theta \left(3 \mu^2_A - 4 \pi^2 T^2 \right) &= 0 \, , \\
\left(4 \pi^2 T^2 - \mu^2_A \right) u^\alpha \partial_\alpha T &= 0  \, ,\\ 
\end{split}
\end{align}
the five vector equations 
\begin{align}
\begin{split}\label{case1eq2}
\left(  4 \pi^2 T^2 - \mu^2_A\right) a^\mu &= 0 \, , \\
\left(4 \pi^2 T^2 - \mu^2_A \right) \Delta^{\mu \nu} \partial_\nu T&= 0 \, , \\
\left( 4 \pi^2 T^2 -\mu^2_A \right) \left( b_I - c_I \right) v^I_\nu &= 0 \, , \\
4 \pi^2 T^2 \mu_A \omega_\nu - \mu^2_A c_I v^I_\nu  &= 0 \, , \\
d_I t^I_{\alpha \beta} \epsilon^{\mu \nu \alpha \beta} u_\nu + a^\mu \left(4 \pi^2 T^2 - 5 \mu_A^2 \right) &= 0 \, , 
\end{split}
\end{align}
and the two tensor equations 
\begin{align}
\begin{split}\label{case1eq3}
\mu_A \epsilon_{\mu \nu \alpha \beta} u^\alpha a^\beta &= 0 \, , \\
\mu_A \epsilon_{\mu \nu \alpha \beta} u^\alpha \partial^\beta T &= 0 \, .
\end{split}
\end{align}
The generic solution to these equations \eqref{case1eq1}-\eqref{case1eq3} --- assuming $\mu_A$ independent of the temperature ---  is obtained by setting the metric coefficients 
\begin{align}
\begin{split}\label{case1IntegrationC}
b_I &= c_I \, , \\
c_I v^I_\nu &= \frac{4 \pi^2 T^2}{\mu_A} \omega_\nu \, , \\
d_I t^I_{\mu \nu} &= 0 \, , \\ 
\end{split}
\end{align}
and imposing the following requirements on the hydrodynamic flow and the spin source $\mu_A$: 
\begin{align}
\begin{split}\label{case1Solution}
\Theta &= 0 \, ,\\
a^\mu &= 0 \, , \\
\partial_\mu T &= 0 \, , \\
u^\alpha \partial_\alpha \mu_A &= 0 \, . \\ 
\end{split}
\end{align}
We can now use equations \eqref{case1IntegrationC} and \eqref{case1Solution} to determine the energy momentum tensor and the spin current as
\begin{align}
\begin{split}\label{energyCase1}
T^{\mu \nu}&=  32 \pi^4 \kappa T^4 \left[ \left(1 - \frac{\mu^2_A}{2 \pi^2 T^2} \right) \left(4 u^\mu u^\nu + \gamma^{\mu \nu}\right)- \frac{4 u^{(\mu} \omega^{\nu)}}{\mu_A} +\frac{2 \mu_A \omega^\mu u^\nu - \epsilon^{\mu \nu \alpha \beta} u_\alpha \partial_\beta \mu_A}{2\pi^2 T^2}  \right] \, ,
\end{split} 
\end{align}
\begin{align}
\begin{split}\label{spinCase1}
S^{\lambda \mu \nu} &= 32 \pi^2 \kappa T^2 \epsilon^{\lambda \mu \nu \alpha} \left( \mu_A u_\alpha + \omega_\alpha  \right) \, .
\end{split}
\end{align}
Separate components of the energy momentum tensor are, see (\ref{compTmn}), 
\begin{align}
\begin{split}\label{hydroProj1}
\varepsilon &= 96 \pi^4 \kappa T^4 \left(1 - \frac{\mu^2_A}{2 \pi^2 T^2} \right) \, , \\ 
p &= -32 \pi^2 \kappa T^4 \left( 1 - \frac{\mu^2_A}{2 \pi^2 T^2} \right) \, , \\
\bar q^\mu &=- \frac{64 \pi^4 \kappa T^4}{\mu_A}   \omega^\mu \, , \\
q^\mu &=- \frac{64 \pi^4 \kappa T^4}{\mu_A} \omega^\mu  \left(1 - \frac{\mu^2_A}{2 \pi^2 T^2} \right)  \, , \\
\pi^{\mu \nu} &= 0 \, ,\\
\tau^{\mu \nu} &=- 16 \pi^2 \kappa T^2 \epsilon^{\mu \nu \alpha \beta} u_\alpha \partial_\beta \mu_A \, .
\end{split}
\end{align}
We note that the energy-momentum tensor would be symmetric only in the limit $\mu_A\to 0$ but this limit is singular as we assumed it non-vanishing in deriving (\ref{energyCase1}), therefore it is generically non-symmetric. On the other hand it is always traceless because the intrinsic torque is antisymmetric in the indices and the vorticity is perpendicular to the fluid velocity. This is as required from a conformal fluid.  The intrinsic torque $\tau^{\mu \nu}$ may vanish if the chemical potential $\mu_A$ is constant in directions tangent to the fluid velocity.\footnote{Just as in regular hydrodynamics, we expect \eqref{hydroProj1} to be frame dependent. This particular frame is automatically determined by our choice of counter terms in the action.} 

The spin current (\ref{spinCase1}) only contains a single irreducible axial current
\begin{align}\label{axialCurrentCase1}
J_A &= -32 \pi^2 \kappa T^2 \left(\mu_A u^\mu + \omega^\mu \right) \, .
\end{align}
This axial current \eqref{axialCurrentCase1} comprises of a charge density $\rho_A = 32 \pi^2 \kappa T^2 \mu_A$ and a linear response proportional to the vorticity with coefficient $32 \pi^2 \kappa T^2$. This last contribution has the same form as the known chiral separation vortical effect (CVSE) \cite{SonSurowka,LandsteinerReview} that is typically associated to anomalous transport in chiral fluids. We note however it resembles more the chiral torsional effect \cite{ChiralTorsional} where an axial current is generated due to the presence of defects. The energy currents in \eqref{hydroProj1} are also akin to CVSE.  However in the latter there exist a single energy current and the proportionality constant has a different dependence on $\mu_A$ and $T$. These differences do not imply any contradiction as the axial component of the spin current in general is different than the axial charge current. All in all, the appearance of vorticity as a source of the spin current in (\ref{spinCase1}) is interesting, implying magnetization by rotation, akin to the Barnett effect \cite{PhysRev.6.239}. 

\subsection{Non vanishing $\mathcal{V}^\mu_2$}

Next, we work out an example with a non-vanishing vector-like spin source $\mathcal{V}^\mu_2$.  The constraint equations in this case can be split int  four scalar equations
\begin{align}
\begin{split}\label{case1Eq1}
\left(12 \pi^2 T^2 - \mathcal{V}^2_2 \right) u^\alpha \partial_\alpha T &= 0 \, , \\
\left(4 \pi^2 T^2 - \mathcal{V}^2_2 \right) \Theta - 2 u^\alpha \mathcal{V}^\beta_2 \partial_\alpha \mathcal{V}_{2 \beta} &=0 \, ,\\
\left(12 \pi^2 T^2 - \mathcal{V}^2_2 \right) \Theta - 3 \mathcal{V}^\alpha_2 \mathcal{V}^\beta_2 \sigma_{\alpha \beta} - 3 c_I v^I_\alpha \mathcal{V}^\alpha_2 &=0 \, , \\
d_I t^I_{\alpha \beta} \mathcal{V}^\alpha_2 \mathcal{V}^\beta_2 &= 0 \, ,
\end{split}
\end{align}
five vector equations 
\begin{align}
\begin{split}\label{case1Eq2}
2 \left( 4 \pi^2 T^2 - \mathcal{V}^2_2 \right) a_\mu + d_I t^I_{\mu \alpha} \mathcal{V}^\alpha_2  &= 0  \, ,\\
4 \pi^2 T^2 a^\mu - \mathcal{V}^\alpha_2 a_\alpha \mathcal{V}^\mu_2 &= 0  \, , \\
4 \pi^2 T \Delta^{\alpha \mu}\partial_\alpha T -  \mathcal{V}^\mu_{2}\mathcal{V}^\alpha_2 \partial_\alpha T &=0  \, ,\\
\left(4 \pi^2 T^2 \gamma^{\mu \alpha} - \mathcal{V}_2^\mu \mathcal{V}_2^\alpha  \right) \left(b_I - c_I\right) v^I_\alpha  &= 0  \, ,\\
\epsilon_{\mu \alpha \beta \rho} u^\alpha \mathcal{V}^\beta_2 \omega^\rho - \frac{4}{3} \mathcal{V}_{2 \mu} \Theta + 2 \mathcal{V}^\alpha_2 \sigma_{\mu \alpha} + \frac{ \mathcal{V}^\alpha_2 \mathcal{V}_{2 \mu} + \mathcal{V}^2_2 \delta^\alpha_\mu }{4 \pi^2 T^2} c_I v^I_{\alpha} &= 0  \, ,
\end{split}
\end{align}
and two tensor equations 
\begin{align}
\begin{split}\label{case1Eq3}
a^{[\mu} \mathcal{V}^{\nu]}_2 &= 0  \, ,\\
\mathcal{V}^{[\mu} \Delta^{\nu]\alpha} \partial_\alpha T &= 0 \, .
\end{split}
\end{align}
See (\ref{uDecomp}) for the definition of the flow parameters. The generic solution to \eqref{case1eq1}-\eqref{case1eq3} with $\mathcal{V}^\mu_2$ independent of temperature is given by the following metric coefficients 
\begin{align}
\begin{split}\label{case2IntegrationC}
b_I &= c_I \, ,\\
c_I v^I_\nu &= \frac{4 \pi^2 T^2  \left( \Theta \mathcal{V}_{2 \nu} - \epsilon_{\nu \alpha \beta \rho} u^\alpha \mathcal{V}^\beta_2 \omega^\rho \right) }{\mathcal{V}^2_2} \, ,\\
t^I_{\mu \nu} &= 0  \, ,
\end{split}
\end{align}
together with the following flow parameters:
\begin{align}
\begin{split}\label{case2Solution}
\Theta &= \frac{2 u^\alpha \mathcal{V}^\beta_2 \partial_\alpha \mathcal{V}_{2 \beta}}{4 \pi^2 T^2 - \mathcal{V}^2_2} \, , \\
\sigma^{\mu \nu} &= \frac{\Theta}{6} \left( \Delta^{\mu \nu} - \frac{3 \mathcal{V}^\mu_2 \mathcal{V}^\nu_2}{\mathcal{V}^2_2}  \right) \, ,\\
a^\mu &= 0 \, , \\
\partial_\alpha T &= 0 \, . 
\end{split}
\end{align}
We note that whenever the vector source $\mathcal{V}_2^\mu$ is time independent, $u^\alpha \partial_\alpha  \mathcal{V}_2^\mu=0$, both the compressibility and the shear tensor vanish. Using \eqref{case2IntegrationC} and \eqref{case2Solution} we arrive at the following energy momentum tensor
\begin{align}
\begin{split}\label{energyCase2}
T^{\mu \nu} &=  32 \pi^4 \kappa T^4 \left[    4 u^\mu u^\nu + \gamma^{\mu \nu} - \frac{ \mathcal{V}^2_2 u^\mu u^\nu + \mathcal{V}^\mu_2 \mathcal{V}^\nu_2}{2 \pi^2 T^2} + \left( \frac{u^\mu \epsilon^\nu_{\hphantom{\nu} \alpha \beta \rho} }{ 4 \pi^2 T^2} +  \frac{ u^{(\mu} \epsilon^{\nu)}_{\hphantom{\nu} \alpha \beta \rho} }{\mathcal{V}^2_2} \right) u^\alpha \mathcal{V}^\beta_2 \omega^\rho \right. \\ &\hphantom{=} \hphantom{-32 \pi^4 \kappa}  \left. + \left( \frac{ \mathcal{V}^\mu_2 u^\nu + 2 u^\mu \mathcal{V}^\nu_2}{2 \pi^2 T^2} - \frac{4 u^{(\mu} \mathcal{V}^{\nu)}_2}{\mathcal{V}^2_2} \right) \Theta  + \left(\frac{ \Delta^{ \mu [\alpha}\Delta^{\nu]}_\beta + 2  u^\mu u^{[\alpha} \Delta^{\nu]}_\beta }{\pi^2 T^2}  \right) \partial_\alpha \mathcal{V}^\beta_2   \right]\, ,
\end{split}
\end{align}
with the following hydrodynamic projections 
\begin{align}
\begin{split}\label{energyProjCase2}
\epsilon &= 96 \pi^2 \kappa T^2 \left(1 - \frac{\mathcal{V}^2_2 + 2 \partial_\alpha \mathcal{V}^\alpha_2}{6 \pi^2 T^2} \right) \, , \\
p &= -32 \pi^2 \kappa T^2 \left(1 - \frac{\mathcal{V}^2_2 + 2 \partial_\alpha \mathcal{V}^\alpha_2}{6 \pi^2 T^2} \right) \, ,  \\
\bar q^\mu &= -\frac{ 4 \pi^2 T^2 - \mathcal{V}^2_2 }{4 \pi^2 T^2 \mathcal{V}^2_2} \Theta \mathcal{V}^\mu_2 - \frac{\epsilon^{\mu}_{\hphantom{\mu} \nu \rho \sigma} u^\nu \mathcal{V}^\rho_2 \omega^\sigma}{\mathcal{V}^2_2} \, ,\\
q^\mu &= -\frac{2 \pi^2 T^2 - \mathcal{V}^2_2}{2 \pi^2 T^2 \mathcal{V}^2_2} \Theta \mathcal{V}^\mu_2 - \frac{\left(8 \pi^2 T^2 + \mathcal{V}^2_2 \right) \epsilon^{\mu}_{\hphantom{\mu} \nu \rho \sigma} u^\nu \mathcal{V}^\rho_2  \omega^\sigma }{8 \pi^2 T^2 \mathcal{V}^2_2} - \frac{u^\alpha \partial_\alpha \mathcal{V}^\mu_2}{2 \pi^2 T^2} \, , \\
\pi^{\mu \nu} &= - 16 \pi^2 \kappa T^2 \left( \Delta^\mu_{(\rho} \Delta^\nu_{\sigma)} - \frac{1}{3} \Delta^{\mu \nu}\Delta_{\rho \sigma} \right) \left( \mathcal{V}^\rho_2 \mathcal{V}^\sigma_2 - \partial^\rho \mathcal{V}^\sigma_2 \right) \, ,\\
\tau^{\mu \nu} &= 16 \pi^2 \kappa T^2 \Delta^\mu_{[\rho} \Delta^\nu_{\sigma]} \partial^\rho \mathcal{V}^\sigma_2 \, .
\end{split}
\end{align}
A number of observations are in order. First, we observe that the energy momentum tensor is not symmetric but it remains traceless as required from a conformal fluid. Then, we see two types of heat currents $q$ and $\bar q$ with components along the spin source $\mathcal{V}^\mu_2$ and along a direction orthogonal to both $\mathcal{V}^\mu_2$ and the vorticity $\omega^\nu$. The shear tensor and the intrinsic torque are completely determined by the symmetric traceless and antisymmetric projections of the gradient of the vector source. 

An interesting specific case is when the source $\mathcal{V}^\mu_2$ is constant\footnote{A simpler case would be vanishing $\mathcal{V}^\mu_2$, but this is not well defined as we assumed it non-vanishing in the derivation.}. In this case the fluid is required to be incompressible, see (\ref{case2Solution}). As a result, several terms in (\ref{energyProjCase2}) vanish, including the first order contribution to the shear tensor but, interestingly, the heat currents remain and given by the second terms in $q$ and $\bar q$ alone. 

The spin current follows from \eqref{case2IntegrationC} and \eqref{case2Solution} as
\begin{align}
\begin{split}
S^{\lambda \mu \nu} &=  32 \pi^2 \kappa T^2 \left[ \Delta^{\lambda [\mu} \mathcal{V}^{\nu]}_2 +2 u^\lambda u^{[\mu} \mathcal{V}^{\nu]}_2 - \Delta^{\lambda [\mu} u^{\nu]} \Theta + \frac{\mathcal{V}^\lambda_2 \mathcal{V}^{[\mu} u^{\nu] }\Theta }{\mathcal{V}^2_2 }  \right. \\ & \hphantom{=} \left. \hphantom{-32 \pi^2 \kappa T^2} + \frac{ \epsilon_{ \alpha \beta \rho \sigma} u^\alpha \mathcal{V}_{2}^{\beta} \omega^\sigma \left(\gamma^{\rho \lambda} u^{[\mu} \mathcal{V}_2^{\nu]} + u^\lambda \gamma^{\rho [\mu} \mathcal{V}^{\nu]}_2  \right)  }{ \mathcal{V}^2_2 } \right] \, ,
\end{split}
\end{align}
with the  following hydrodynamic projections 
\begin{align}
\begin{split}\label{projSV}
\rho_V &=  \frac{32}{3} \pi^2 \kappa T^2 \Theta \, , \\
\rho_A &= 0 \, ,\\
n^\mu_V &= -32 \pi^2 \kappa T^2 \mathcal{V}^\mu_2 \, ,\\
\bar n^\mu_V &=  16 \pi^2 \kappa T^2 \mathcal{V}^\mu_2 \, ,\\
n^\mu_A &= -16 \pi^2 \kappa T^2 \left( \gamma^{\mu \nu} - \frac{\mathcal{V}^\mu_2 \mathcal{V}^\nu_2}{\mathcal{V}^2_2}   \right) \omega^\nu \, ,\\
\bar n^\mu_A &= 8 \pi^2 \kappa T^2 \left( \gamma^{\mu \nu} - \frac{\mathcal{V}^\mu_2 \mathcal{V}^\nu_2}{\mathcal{V}^2_2}   \right) \omega^\nu \, ,\\
N^\lambda_\kappa &= -4 \pi^2 \kappa T^2 \left( \frac{10}{3} \Delta^\lambda_\kappa    - 6  \frac{\mathcal{V}^\lambda_2 \mathcal{V}_{2 \kappa}}{\mathcal{V}^2_2} \right) \Theta  +\frac{ 8 \pi^2 \kappa T^2   \left(2 \gamma^{\alpha \lambda} \gamma_{\nu \kappa} + \delta^\lambda_\nu \delta^\alpha_\kappa  \right)\epsilon_{\alpha \beta \rho \sigma} \mathcal{V}^\nu_2 u^\beta \mathcal{V}^\rho_2 \omega^\sigma}{\mathcal{V}^2_2} \, , \\
\bar N^\lambda_\kappa &= 0 \, .
\end{split}
\end{align}
The corresponding vector and axial components of the spin current  are rewritten in terms of the irreducible currents as
\begin{align}
\begin{split}\label{currentsSV}
J^\mu_V &= -\frac{32}{3} \pi^2 T^2 \Theta u^\mu \, ,\\
\bar J^\mu_V &= -32 \pi^2 \kappa T^2 \mathcal{V}^\mu_2 \, ,\\
J^\mu_A &= 0 \, ,\\
\bar J^\mu_A &= -16 \pi^2 \kappa T^2 \left( \gamma^{\mu \nu} - \frac{\mathcal{V}^\mu_2 \mathcal{V}^\nu_2}{\mathcal{V}^2_2}   \right) \omega_\nu \, .
\end{split}
\end{align}
Perhaps the most remarkable finding in this analysis is the last equation:  a novel type of torsional anomalous transport in the direction proportional to the projection of vorticity transverse to the spin source vector. We leave a detailed study of spin induced anomalous transport like this to  future work.

\section{Discussion}
\label{sec::Conclusions}

In this work we considered strongly interacting relativistic quantum field theories with spin degrees of freedom in the hydrodynamic limit and established the constitutive relations of the basic hydrodynamic variables, the energy-momentum tensor and the spin current based on the classification of the spin sources in irreducible Lorentz representations. This decomposition of the sources and the hydrodynamic variables is summarized in table \ref{table}. 
A general conclusion is the possibility of generating non-trivial spin current in flat space-time, even in the case where the energy-momentum tensor is symmetric, by inclusion of non-trivial torsion. 
 
To find specific examples of hydrodynamics with spin current, we specified to 3+1 dimensional conformal fluids and calculated the components of the aforementioned hydrodynamic variables at the first two orders in the derivative expansion using holographic methods In particular we assumed the fluid to have dual gravitational description as a blackhole solution in 5D Lovelock-Chern-Simons theory. The reason for this non-standard choice of holography is twofold: (i) the need to go beyond Einstein's gravity to keep the vielbein and spin connection as independent variables which is required to obtain a non-vanishing spin current in flat space-time, (ii) the simplicity of this theory which allowed us to construct the hydrodynamic flow analytically (in the derivative expansion), by reducing the gravitational equations of motion to algebraic constraints. Clearly, we do not expect this holographic theory to represent the spin liquids found in Nature, but this is not the point of the paper. The holographic theory we consider should be viewed as a scaffold to help solve the hydrodynamic flow equations. Any such analytic solution is instructive and valuable. 

Two of our particularly interesting findings are (i) existence of a dynamical version of the Barnett effect where vorticity of the liquid generates a spin current. This can also be viewed as some sort of anomalous transport analogous to the chiral vortical separation effect where an axial current is generated in the direction of the vorticity with conductivity proportional to $T^2$. (ii) a novel type of anomalous vortical transport transverse to the spin vector source, a sort of ``spin vortical axial'' Hall effect. It is intriguing to see whether any of this transport phenomena exhibit universality as in the chiral magnetic or vortical effects. 

Apart from these hydrodynamic findings the analytic blackhole solutions that we derived here are novel. ln particular we have found a general class of black hole solutions to 5D Lovelock-Chern-Simons gravity with non-trivial spacetime torsion. These will hopefully be useful beyond holography. 

Our work should be viewed as a first step towards the study of strongly coupled relativistic QFT's and strongly correlated relativistic fluids with non-trivial spin currents. There are various directions to continue: 

\begin{enumerate} 

\item One obvious extension is a systematic study of  hydrodynamics with spin currents and the associated transport properties. This includes listing all linearly independent transport coefficients allowed by the underlying symmetries and working out the constraints that arise from positive entropy generation and Onsager relations. It would be very interesting to include electromagnetic fields in this analysis. 

\item Another possible extension is generalization of our study to generic 5D Lovelock gravity and/or different Chern-Simons gauge choice. Following \cite{Banados} we have explicitly used the Chern-Simons structure to derive the holographic dictionary. We are curious  whether our findings can hold beyond this. For example in a more generic Lovelock type theory 
\begin{equation}
S_{\text{Lovelock}}= \int \epsilon_{ABCDE} \left[c_1 \hat e^A \hat e^B \hat e^C \hat  e^D \hat e^E + c_2 \hat  R^{AB} \hat  e^C \hat e^D \hat e^E + c_3 \hat R^{AB} \hat R^{CD} \hat e^E \right]  \, ,
\end{equation}
with arbitrary coefficients $c_1,c_2$ and $c_3$. Here we do not expect to find generic analytic solutions but the resulting hydrodynamic flow may correspond to more realistic fluids. 

In addition to this, we used a particular gauge was for this dictionary to be applicable to black hole like solutions.  We have considered the most general solution (for independent sources) within this gauge choice, but it remains to be seen if other interesting spin induced hydrodynamic transport can be observed within different gauge choices. 

\item Finally, it is tempting to investigate the effect of the spin sources on the entanglement structure in the quantum field theory, such as entanglement entropy (EE) and the Renyi entropies. This could be done by studying how the Ryu-Takayanagi proposal \cite{Ryu_2006} extends to theories with torsion. Equally interesting is the question whether the Einstein-Cartan equations can also be derived from entanglement laws just as the standard Einstein equations \cite{Jacobson:1995ab,VanRaamsdonk:2010pw,Faulkner:2013ica}.   

\end{enumerate}

\begin{acknowledgements}
This work was partially supported by the Netherlands Organisation for Scientific Research (NWO) under the VIDI grant 680-47-518 and the Delta-Institute for Theoretical Physics (D-ITP), both funded by the Dutch Ministry of Education, Culture and Science(OCW). DG is supported in part by CONACyT through the program Fomento, Desarrollo y Vinculacion de Recursos Humanos de Alto Nivel.
\end{acknowledgements}

\newpage 
\appendix

\section{Boundary Noether Symmetries and Anomalies}
\label{sec::Noether}

The residual gauge transformations that preserve the gauge choice \eqref{gaugeChoiceGeneral} with vanishing $\{H^a_{\pm}, H^{ab} \}$ take the asymptotic form 

\begin{align}
\begin{split}
\tau &= u(x) P_5 + \hat \alpha^a(x)  J^+_a + \hat \beta^a(x) J^-_a + \frac{1}{2}\lambda^{ab}(x) J_{ab} \, , \\
\hat \alpha^a &= \frac{1}{\sqrt{\rho}} \left[ 1 + c_0 \rho^{1/2} + \frac{c_1}{2} \rho + \frac{2 c_2 - c_0 c_1}{6} \rho^{3/2} + \frac{6 c_3 - 4 c_0 c_2 - 3 c_1^2 + 2 c_0^2 c_1}{24} \rho^2 \right] \alpha^a \, , \\
\hat \beta^a &= \sqrt{\rho} \left[1 - c_0 \rho^{1/2} + \frac{2c^2_0- c_1}{2}\rho \right] \beta^a \, ,
\end{split}
\end{align}
with $\{u,\alpha^a,\beta^a,\lambda^{ab} \}$ functions of the boundary coordinates parametrizing the residual boundary symmetries. The corresponding transformation of the bulk fields are 

\begin{align}\label{residualTransformationFields}
\begin{split}
\delta e^5 &= du - 2\left( e^a \beta_a - k^a \alpha_a \right) \, ,\\
\delta e^a &= D\alpha^a - \lambda^a_{\hphantom{a}c} e^c + u e^a + \mathcal{L}_\xi e^a \, ,  \\
\delta k^b &= D\beta^a - \lambda^a_{\hphantom{a}c} k^c - u k^a + \mathcal{L}_\xi k^b \, ,\\
\delta \omega^{ab} &= D\lambda^{ab} + 4 e^{[a} \beta^{b]} + 4 k^{[a} \alpha^{b]} + \mathcal{L}_\xi \omega^{ab} \, ,
\end{split}
\end{align}
where  $\mathcal{L}_\xi$ denotes the Lie derivative with respect to a boundary diffeomorphism\footnote{The gauge condition $N^\nu=0$ ensures that the inclusion of diffeomorphism does not affect the residual gauge analysis. It also implies $\xi^\mu=\xi^\mu(x)$ and in particular we can also absorb $\xi^r$ into the definition of the gauge parameter $u$.} parametrized by  $\xi^\mu=\xi^\mu(x)$ is included for the sake of generality. The covariant derivatives are given by

\begin{align}
\begin{split}
D \alpha^a &= d \alpha^a + \omega^{a}_{\hphantom{a}b} \alpha^b \, , \quad \quad D \beta^a = d \beta^a + \omega^{a}_{\hphantom{a}b} \beta^b\, , \quad \quad D \lambda^{ab} = d \lambda^{ab} + \omega^{a}_{\hphantom{a}c}\lambda^{cb}- \omega^b_{\hphantom{b}c} \lambda^{ca}\, .
\end{split}
\end{align}

\vspace{2mm}
To preserve the block diagonal splitting of the metric we require constraint $\delta e^5=0$ which in turn makes one of  $\{u,\alpha^a,\beta^b \}$ dependent on the others. For definiteness, we consider $\{u,\alpha^a\}$ as the independent ones. 

It is now possible to calculate the classically conserved currents related to these transformations by using \eqref{residualTransformationFields},\eqref{asympSol} and \eqref{fieldOnShell}, obtaining 
\begin{align}
\begin{split}\label{classicalCurrents}
\delta \xi^a  &: \bar A_a \equiv  D\tau_a - \left(I_a T^b \tau_b + \frac{1}{2}I_a R^{bc} \sigma_{bc} \right) + \frac{1}{2} I_a \omega^{cd} \left(D \sigma_{cd} - 2 e_{[a}\tau_{d]} \right)  \overset{\text{\tiny{Classical}}}{=}  0 \, , \\
\delta \lambda_{ab} &: A^{ab} \equiv D\sigma^{ab} - 2 e^{[a}\tau^{b]}  \overset{\text{\tiny{Classical}}}{=}  0 \, ,  \\
\delta u &: A \equiv e^a \tau_a + D\left(e^a I^b \sigma_{ab} \right)  \overset{\text{\tiny{Classical}}}{=}  0  \, , \\
\delta \alpha_a &: A^a  \equiv  D \tau_a - 2\left(e^b \sigma_{b c}e^{c \mu} k_{a \mu} + k^b \sigma_{ba}  \right) \overset{\text{\tiny{Classical}}}{=}  0\, .
\end{split}
\end{align}
The gauge parameters $\{\xi^\mu,\lambda^{ab},u,\alpha^a \}$ parametrize boundary diffeomorphisms, boundary local Lorentz transformations, boundary Weyl transformations and a non-abelian symmetry\footnote{To see that the symmetry should be non-abelian, one calculates the asymptotic group algebra for the transformations \eqref{residualTransformationFields} and observe that the gauge transformation parametrized by $\alpha^a$ induces a non-abelian extension that is non-linearly realized \cite{Miskovic}. For vanishing torsion this transformation is not independent and reduces to a combination of a local diffeomorphism and a local translation.} respectively. For vanishing spin current the energy momentum tensor becomes symmetric and traceless. In this case invariance under diffeomorphism and the non-abelian transformations both yield conservation of energy-momentum. It is now possible to use the equations of motion \eqref{t3} and \eqref{s3} to compute the Noether currents explicitly (up to total derivatives):
\begin{align}
\begin{split}\label{quantumCurrents}
\bar A_a &= 0 \, ,\\
A^{ab} &=0 \, , \\
A &= \kappa \epsilon_{abcd} R^{ab} R^{cd} \, , \\ 
A_a &= 8 \kappa \left(k_a^{\hphantom{a}e} \epsilon_{ebcd} T^b R^{cd} - 4 \epsilon_{abcd} T^b k^c k^d \right)  \, ,
\end{split}
\end{align}
where $k^{ a d} = k^a_\mu e^{\mu d}$. The absence of anomalies in the currents related to local Lorentz tranformations and diffeomorphisms imply the conservation equations \eqref{cons1}. The anomaly in the Weyl current $A$ is the known trace anomaly as the right hand side of A in \eqref{quantumCurrents} is the Euler density. Finally, the non-abelian current turns out to be generically anomalous when the boundary torsion is non vanishing. One remark here is that the non-abelian transformation is nonstandard, as it transforms\footnote{After performing a local lorentz transformation with parameter $\omega^{a b}_\mu \alpha^\mu$ and a diffeomorphism with parameter $\xi^\beta=\alpha^\beta$} the coframes as $\delta e^a_\mu = \alpha^\nu T^a_{\mu \nu} $.

\section{Singular Solutions}\label{sec::Appendix::SingularSolutions}

In section \ref{sec::ZerothSolution} we used the regulatiry of the Ricci scalar $\mathcal{R}$ near the black hole horizon to fix some of the integration constants in \eqref{solutionsGeneral}. Here we relax this requirement and present a solution with only scalar spin sources $\mu_V$ and $\mu_A$ that has divergent $\mathcal{R}$ at the horizon. Working in the same gauge as in \eqref{solutionGeneral0} we find 
\begin{align}
	\begin{split}
		ds^2 &= - \frac{dr^2}{r^2 g^2} + r^2 (-f^2 u_\mu u_\nu + h^2 \Delta_{\mu \nu} )dx^\mu dx^\nu \, , \\
		f^2 &= 1 -\frac{r^2_h}{r^2} \, ,\\
		g^2 &= \left(1 - \frac{r^2_h}{r^2} \right) \left( 1 - \frac{r^2_h - 4 \pi^2 T^2}{r^2} \right) \, ,\\
		h^2 &= \frac{1}{4} \left[  \left(1 - \frac{\mu^2_V - \mu^2_A}{4 \pi^2 T^2} \right) \sqrt{1 -\frac{r^2_h - 4 \pi^2 T^2}{r^2}}  + \left( 1 + \frac{\mu^2_V - \mu^2_A}{4 \pi^2 T} \right) \sqrt{1 - \frac{r^2_h}{r^2}} \right]^2 \, , \\
		T^a &=  r  \theta^a_\mu \left[ - \mu_A h \epsilon^{\mu}_{\hphantom{\mu} \nu \rho \sigma} u^\nu + \mu_V f u_\rho  \Delta^\mu_\sigma     \right]dx^\rho \wedge dx^\sigma   \, ,\\
T^5 &= 0 \, .
	\end{split}
\end{align}
\section{Thermal Entropy in Riemmann-Cartan Spacetimes}
\label{sec::Appendix::Entropy}

An entropy formula and the corresponding first law for blackhole solutions in arbitrary theories of gravity carrying a metric and a Levi-Civita connection was derived in \cite{WaldOriginal} by Lee and Wald. An analogous formula for gravitational theories described in vielbein formalism, albeit with vanishing torsion, was not addressed until much later\footnote{The complicatio arises from the internal degrees of freedom of the vielbein associated to the Lorentz symmetry of the tangent space. See the discussion in \cite{Ent1} for a summary.} in \cite{Ent1,Ent2,Ent3}. In this appendix we derive the blackhole entropy formula for gravitational theories in the first order formalism of gravity, with independent vielbein and connection, by extending the approach in \cite{Ent3}. We first present a quick review of the necessary covariant phase space formalism --- for a more complete review see \cite{covRev1,covRev2,covRev3} --- and then we apply it to gravitational theories that we are interested in. 

\subsection{Covariant phase space formalism}

To start the discussion lets consider a local Lagrangian $\mathcal{L}=\mathcal{L}(\phi)$ depending on canonical fields $\{\phi\}$.  Through the variation of the Lagrangian one obtains not only the equations of motion $E_\phi$ but also the so called symplectic potential $\theta(\delta \phi)$
\begin{align}\label{variationLagrangianCovariant}
\delta \mathcal{L} = E_\phi \delta \phi + d \theta(\delta  \phi ) 
\end{align}
It is important to note that $\theta$ is not uniquely defined as any exact form $d \alpha$ could be added to it. We instead consider a class of symplectic potentials $\Theta$ defined by
\begin{align}
\Theta = \theta + d \alpha 
\end{align}
where the form $\alpha$ will be used to fix the gauge invariance of the symplectic potential. The anti-symmetrized field variation of $\Theta$ defines a symplectic form $\Omega$ as
\begin{align}
\Omega(\delta_1,\delta_2) = \delta_1 \Theta - \delta_2 \Theta \, .
\end{align}
For a symmetry parametrized by $\omega$, the symplectic structure allow us to define a Hamiltonian flow by
\begin{align}
\delta H_\omega = \int_{\Sigma} \Omega \left(\delta, \delta_\omega \right)\, ,
\end{align}
where $\Sigma$ denotes the spacetime manifold. For a diffeomorphism $\xi$ we expect all linear variations $\delta_\xi$ to vanish, as this is a symmetry of the theory, implying that the Hamiltonian flow should also vanish, namely $\delta H_\xi =0$. In Riemann-Cartan spacetimes there is, in addition to diffeomorphisms, local Lorentz symmetry parametrized by $\lambda^{AB}$, so we will equivalently ask for absence of an associated charge by demanding $\delta H_\lambda=0$. This allows us to fix $\alpha$.

\vspace{2mm}
It is also possible to use the symplectic potential to define a Noether current associated to diffeomorphisms $\xi$. This can be done by noting that, if we define this current $J_\xi$ as
\begin{align}
J_\xi = \Theta(\delta_\xi) - \mathcal{L}\cdot \xi \, , 
\end{align}
then $dJ =0$ on shell using \eqref{variationLagrangianCovariant}. This implies we should be able to write the current as an exact form 
\begin{align}
J_\xi = d Q_\xi \, ,
\end{align}
where $Q_\xi$ will be the Noether charge. The Noether charge and the Hamiltonian flow for a diffeomorphism can  easily be related by noticing that
\begin{align}
\Omega (\delta,\delta_\xi) &= d \left(\delta Q_\xi - \Theta(\delta) \cdot \xi \right)\, ,
\end{align}
which implies the Hamiltonian flow is given by a boundary term 
\begin{align}
\delta H_\xi = \int_{\partial \Sigma} \left[\delta Q_\xi - \Theta(\delta) \cdot \xi \right]\, ,
\end{align}
whenever $\Theta \cdot \xi = \delta B$ with $B$ being some differential form.  We then say the theory is integrable. We are concerned with theories where $\partial \Sigma$ is formed by a Killing horizon, i.e. $\xi=0$ on this surface, and with an asymptotic boundary at $\infty$. Taking this into account together with the vanishing of the Hamiltonian flow we are left with 
\begin{align}\label{FirstLawBlackHoles}
\int_{\partial \Sigma_h} \delta Q^h_\xi = \int_{\partial \Sigma_\infty} \left[ \delta Q^\infty_\xi - \Theta^\infty(\delta)\cdot \xi \right] \, .
\end{align}
This equation is the first law of thermodynamics when the left hand side identified with $T_0 \delta S_{\text{thermal}}$. We now proceed to calculate the relevant Noether charge.

\subsection{Gauge Invariant symplectic potential}

To compute $Q_\xi$ it is necessary to find a form $\alpha$ such that $\delta H_\lambda =0$. As in \cite{Ent3} we rather  proceed via a simpler road by requiring  for the symplectic potential itself to be invariant, namely $\Theta(\delta_\lambda)=0$. For completeness, we show below the relevant local Lorentz transformations for the coframes and connection \footnote{Note that at this point we are working in arbitrary dimensions and $A$ represent D-dimensional indices, not necessarily the 5D ones.} 
\begin{align}
\begin{split}
\delta_\lambda e^A &= - \lambda^A_{\hphantom{A} F} \hat e^F\, , \\
\delta_\lambda \omega^{AB} &= D \lambda^{AB}\, ,
\end{split}
\end{align}
We now consider gravitational Lagrangians that are functions of both the coframes $e^A$ and the connection $\omega^{AB}$. In particular we expect the Lagrangian to depend on the local Lorentz covariant forms $\{e^A, T^A, R^{AB} \}$, i.e. $\mathcal{L} = \mathcal{L}\left(e ,T ,R \right)$.

\vspace{2mm}
The corresponding variation of the Lagrangian can then be written as
\begin{align}
\delta \mathcal{L} = \delta e^A E_A + \delta \omega^{AB} E_{AB} + d \Theta(\delta e, \delta \omega )\, ,
\end{align}
 with $E_A$, $E_{AB}$ the equations of motion and $\Theta$ the symplectic potential given by 
 \begin{align}
 \begin{split}
 E_A &= \frac{\partial \mathcal{L}}{\partial e^A} + D \left( \frac{\partial \mathcal{L}}{\partial T^A} \right) \, ,\\
 E_{AB} &= e_B \frac{\partial \mathcal{L}}{\partial T^A} + D \left( \frac{\partial \mathcal{L}}{\partial R^{AB}} \right) \, ,\\
 \Theta(\delta e, \delta \omega ) &= \delta e^A \frac{\partial \mathcal{L}}{\partial T^A} + \delta \omega^{AB} \frac{\partial \mathcal{L}}{\partial R^{AB}} + d \alpha \, .
 \end{split}
 \end{align}
We obtain
\begin{align}
\begin{split}
\int_{\Sigma} \Theta \left( \delta_\lambda \right) &= \int_{\Sigma} \left[ -\lambda^A_{\hphantom{A} F} e^F \frac{\partial \mathcal{L}}{\partial T^A}  + D \lambda^{AB} \frac{\partial \mathcal{L}}{\partial R^{AB}} + d \alpha(\delta_\lambda) \right] \\
&= -\int_{\Sigma} \lambda^{AB} \left[e_B \frac{\partial{\mathcal{L}}}{\partial T^A} + D \left( \frac{\partial \mathcal{L}}{\partial R^{AB}} \right) \right] + \int_{\partial_{\Sigma}} \left[ \lambda^{AB} \frac{\partial \mathcal{L}}{\partial R^{AB}} + \alpha(\delta_\lambda) \right] \\
&= - \int_{\sigma} \lambda^{AB} E_{AB} + \int_{\partial_\Sigma} \left[ \lambda^{AB} \frac{\partial \mathcal{L}}{\partial R^{AB}} + \alpha(\delta_\lambda)  \right]\, .
\end{split}
\end{align}
Requiring  on-shell gauge invariance of the potential fixes $\alpha$ as
\begin{align}
\alpha(\delta) = - \left( e^{a \mu} \delta e^b_\mu \right) \frac{\partial \mathcal{L}}{\partial R^{ab}}\, .
\end{align}
We note that the form of $\alpha$ is valid for any theory in a Riemann-Cartan spacetime. To calculate the Noether we will need to specialize to a particular Lagrangian.

\subsection{Noether Charge in 5D Lovelock Gravity}

We consider a (non necessarily Chern-Simons) 5D Lovelock Lagrangian characterized by the action 
\begin{align}
S =   \int \epsilon_{ABCDF} \left[ c_1 R^{AB} R^{CD} e^F + \frac{c_2}{3} R^{AB} e^C e^D e^F + \frac{c_3}{5} e^A e^B e^C e^D e^F\right] \, ,
\end{align}
with free parameters $\{c_1,c_2,c_3\}$. The equations of motion for this action read
\begin{align}
\begin{split}
E_A = \epsilon_{ABCDF} \left[c_1 R^{BC} R^{DF} + c_2 R^{BC} e^D e^F + c_3 e^B e^C e^D e^F \right] \\
E_{AB}= \epsilon_{ABCDF} \left[ 2 c_1 R^{CD} + c_2 e^C e^D \right] T^F \, , 
\end{split}
\end{align}
with the symplectic potential $\Theta$
\begin{align}
\Theta = \epsilon_{ABCDF} \left( 2 c_1 R^{AB} e^C + \frac{c_2}{3} e^A e^B e^C \right) \delta \omega^{DF}\, ,
\end{align}
and the symplectic current
\begin{align}
J = d \left[ \epsilon_{ABCDF} \left(2c_1 R^{AB} + \frac{c_2}{3} e^A e^B e^C \right)\left( e^{D \mu} T^{F}_{\mu \nu} \xi^\nu + D^{D} \xi^F  \right) \right] + E^{AB} \left( \omega_{AB} \cdot \xi   \right) + E^A \left( e_A \cdot \xi  \right)\, .
\end{align}
We can now read off the Noether charge and evaluate it at the horizon, by noting that $D^{[F} \xi^{F]} \rightarrow n^{DF}$. This yields the following expression for the entropy 
\begin{align}
\label{WaldEntropy}
S = 2 \pi \int_{M_3} \left[ n^{AB} \frac{\partial \mathcal{L}}{\partial R^{AB}} \right] = 2 \pi \int_{M_3} \epsilon_{ABCDF}n^{AB} \left[ 2c_1 R^{AB} + \frac{c_2}{3} e^A e^B e^C \right] \, .
\end{align}
with the $2 \pi$ a normalization factor. Setting $c_1=1$ and $c_2 =2$ we find the result of \eqref{entropyCS}.

\bibliographystyle{ieeetr} 
\bibliography{bibliography2}

\end{document}